\documentclass[a4paper,12pt]{article}
\pdfoutput=1
\usepackage{amsmath}
\usepackage{amsfonts}
\usepackage{amssymb}
\usepackage{amstext}
\usepackage{graphicx}
\usepackage{xcolor}
\newcommand{\ti}[1]{\ensuremath{ \tilde{#1}}}
\newcommand{\nb}{\ensuremath{ \nabla }}

\newcommand{\m}{\ensuremath{{\mu \nu}}}

\newcommand{\p}{\ensuremath{\partial{}}}
\newcommand{\fx}{\ensuremath{\dfrac{x^4}{4}-\dfrac{x^2}{2} }} 
\usepackage{epsfig}%
\title{\centerline \bf Dynamical stability of $k$-essence field interacting non-minimally with a perfect fluid}
\bigskip
\author{Anirban Chatterjee$^\dagger$, Saddam Hussain$^\ddagger$,Kaushik Bhattacharya$^\$$
\thanks{$^\dagger$anirbanc@iitk.ac.in, $^\ddagger$msaddam@iitk.ac.in, $^\$$kaushikb@iitk.ac.in}
\\
\normalsize
Department of Physics, Indian Institute of Technology, Kanpur\\ 
\normalsize
Kanpur 208016, India
}
\begin{document}
\maketitle
\begin{abstract}
We study models of non-minimally coupled relativistic fluid and $k$-essence scalar field in the background of a flat Friedmann-Lemaitre-Robertson-Walker universe.  The non-minimal coupling term is introduced in the Lagrangian level. We employ the variational approach with respect to independent variables that produce modified $k-$essence field equations and the Friedmann equations. We have analyzed the coupled field-fluid framework explicitly using the dynamical system technique considering two different models based on inverse power-law potential. After examining these models it is seen that both models are capable of producing accelerating attractor solutions  satisfying adiabatic sound speed conditions. 
\end{abstract}
\section{Introduction}
\label{sec:intro}
	
Over the past few decades, several cosmological observations have established that our universe has undergone a transition from a decelerated phase to an accelerated phase of expansion in our cosmic evolution. Measurements of luminosity distances and red-shifts of Type Ia supernova \cite{Riess:1998cb,Perlmutter:1998np,Riess:2006fw} are primarily used to support this fact. Other independent observations like Baryon Acoustic Oscillations \cite{Eisenstein:2005su, Percival:2006gs}, Cosmic Microwave Background radiations \cite{Gawiser:2000az}, the power spectrum of matter distributions have also reinforced this idea of a late-time cosmic acceleration. Theoretically the cause of this late time cosmic acceleration is attributed to a hypothetical unclustered form of energy present in the universe, traditionally called  `Dark Energy.' Cosmic acceleration is driven by the negative pressure of the dark energy component. On the other hand, observations of rotation curves of spiral galaxies \cite{Sofue:2000jx}, gravitational lensing\cite{Bartelmann:1999yn}, Bullet cluster, and other colliding clusters give support in favor of the existence of non-luminous matter in the universe. Such matter primarily interacts through gravitational interactions and it is distinct from the baryonic matter and  hence  termed as `Dark Matter.'  WMAP \cite{Hinshaw:2008kr} and Planck \cite{Ade:2013zuv} experiments established the fact that at present epoch, $ 96\% $ of total energy density of the universe is due to dark matter ($\sim 27 \%$) and dark energy ($\sim 69 \%$) contributions. The rest of the $ 4\% $ contribution comes from the  Baryonic matter and radiation part.
	
Diverse theoretical approaches are used to construct different models for dark energy\cite{DES:2021wwk} to explain the late-time cosmic acceleration. One kind of such dark energy model involves field-theoretic models where the dynamics of a scalar field is usually exploited to generate a negative pressure resulting in late-time cosmic acceleration. Depending on the form of the Lagrangians, two major class of models viz. `Quintessence' models \cite{Peccei:1987mm,Ford:1987de,Peebles:2002gy,Nishioka:1992sg, Ferreira:1997au,Ferreira:1997hj,Caldwell:1997ii,Carroll:1998zi,Copeland:1997et,Zlatev:1998tr,Hebecker:2000au,Hebecker:2000zb} and `$k$-essence' models \cite{Fang:2014qga,ArmendarizPicon:1999rj,ArmendarizPicon:2000ah,ArmendarizPicon:2000dh,ArmendarizPicon:2005nz,Chiba:1999ka,ArkaniHamed:2003uy,Caldwell:1999ew}  have been widely studied and discussed. There exist diverse theoretical models of dark matter to explain the present-day cosmic acceleration. The cold dark matter model is one of them.  This model produces fine agreement with several cosmological observations. However, one of the serious disadvantages of this model is the accidental coincidence of magnitudes of observed values of dark matter and dark energy densities of the universe at the present epoch. Out of the several theoretical approaches used to resolve this issue, one engrossing technique is the study of interacting field-fluid scenarios. Various field-fluid coupling scenarios have already been investigated in literature \cite{Brown:1992kc,Boehmer:2015kta,Boehmer:2015sha,Shahalam:2017fqt,Barros:2019rdv,Kerachian:2019tar}. In most of the cases the field sector is governed by the `Quintessence' scalar field and the fluid part is composed of a pressure-less perfect fluid. The `Quintessence' scalar field and perfect fluid can have non-minimal  algebraic \cite{Boehmer:2015kta} or derivative \cite{Boehmer:2015sha} couplings. 

In this paper we have advanced the approach that was initiated in \cite{Boehmer:2015kta}. We have studied a model where the $k-$essence scalar filed is non-minimally coupled to a perfect relativistic fluid. Traditionally the $k-$essence sector is capable of producing accelerated expansion of the universe and the pressure-less fluid can act as dark matter constituent. In our work both these sectors interact with each other and exchange energy and momentum. In this interacting field-fluid model the dark energy sector and dark matter sector acts in a combined way as the dark sector. The interaction term is introduced at the Lagrangian level. Here the relativistic perfect fluid is specifically chosen for the pressure-less dark matter fluid. We have introduced the interaction term which depends on the number density ($n$) and entropy density ($s$) of the fluid constituents, the scalar field $\phi$, and the kinetic term $X$ of the $k-$essence field.  We call it mixed (between algebraic and derivative) coupling because of the presence of derivative of $\phi$ in $X$. We have employed the variational approach to obtain the coupled equations of motion of dark matter and dark energy in isotropic and homogeneous Friedmann-Lemaitre-Robertson-Walker (FLRW) background.  After setting up the basic cosmological equations for this field-fluid model we have further studied the dynamical stability of the system by introducing different kinds of interaction models in the presence of inverse square law potential $V(\phi) \propto {1}/{\phi^2} $\cite{Yang:2010vv}. The dynamical system technique turns out to be a very useful tool to understand the evolution of cosmological models. The techniques we have used relies on previous works in this field \cite{Boehmer:2015sha,Pal:2019tkc,Chakraborty:2019swx,Roy:2017uvr,DeSantiago:2012nk,Dutta:2016bbs,Ng:2001hs,Koivisto:2009fb,Bahamonde:2017ize,Tamanini:2014mpa,Dutta:2017wfd} which were done earlier to study different physical systems. Our work shows that the coupled dark sector can produce accelerating universe solution in the late cosmological era. Some of the solutions are well behaved and shows that our model can genuinely be taken as an alternative to $\Lambda$CDM models. One of the interesting features of our work is related to the fact that the non-minimally interacting field-fluid model can effectively resemble radiation phase as well, although we have not explicitly used any radiation component in our analysis. This fact can be useful when one tries to join the late time cosmological development with the earlier radiation dominated phase.

The paper is organized as follows. We have employed the variational approach to derive the field equation and the expression of stress energy tensor in section~[\ref{sec:2}]. In section~[\ref{sec:3}] we have discussed field-fluid interaction theory in the context of a flat FLRW universe. In section~[\ref{sec:4}]  we give an explicit overview on dynamical stability theory and discuss various outcomes of our model.  Finally we conclude in section~[\ref{sec:5}].
\section{Interaction of relativistic fluid and $k$-essence scalar field: Non-minimal coupling}
\label{sec:2}

The idea of $k$-essence theory was first introduced by Armendariz-Picon et al. \cite{ArmendarizPicon:1999rj, ArmendarizPicon:2000ah}  to explain the inflationary scenario of the early universe, later this theory has also been applied to study the late time phase of cosmic evolution. $k$-essence theory is characterized by a scalar field $\phi$ with a non-canonical kinetic term $X$. Kinetic term can be written in terms of derivatives of $\phi$ as, $ X=-\frac12 (\nb^\mu \phi)(\nb_\mu \phi)\, $. In our model, scalar field $\phi$ interacts with a matter fluid \cite{Brown:1992kc} through the non-minimal coupling term in the Lagrangian. We have studied this field-fluid coupling at the Lagrangian level by employing the variational method \cite{Boehmer:2015kta}. The grand action can be written for this field-fluid theory as:
%
%
\begin{eqnarray}
S&=&\int_\Omega d^4x \left[\sqrt{-g}\frac{R}{2\kappa^2}-\sqrt{-g}\rho(n,s) +
J^\mu(\varphi_{,\mu} + s\theta_{,\mu} + \beta_A\alpha^A_{,\mu})    
-\sqrt{-g}{\mathcal L}(\phi,X)\right.\nonumber\\
& &\left.-\sqrt{-g}f(n,s,\phi,X)\right]\,.
\label{act}
\end{eqnarray}
Here $\kappa^2=8\pi G$. The first term on the right hand side specifies the gravitational part of the action. The next two terms describe the action for a perfect fluid. The third term represents the action for the $k$-essence field and the last term specifies the action for non-minimal field-fluid coupling. Here $f(n,s,\phi,X)$ is an arbitrary function of its variables. The functional form of $f$ depends on the choice of our model of non-minimal interaction. We are working with a  mixed type of coupling where the fluid parameters are coupled to the field $\phi$ as well as derivatives of the field appearing in $X$. In general the commas followed by Greek indices, in the subscripts, specify covariant derivatives with respect to $g_{\mu \nu}$. For scalar functions they become ordinary partial derivatives. 

In the fluid action $\rho(n,s)$ stands for the energy density of the fluid which depends on particle number density ($n$) and the entropy density  per particle ($s$). The other variables as $\varphi, \theta$ and $\beta_A$ are all Lagrangian multipliers. Here Greek-indices run from $0$ to $3$ and $\alpha^A$ is a Lagrangian coordinate of the fluid where $A$ takes the values $1,2,3$. The current density $J^{\mu}$ is related to $n$ as 
\begin{equation}\label{}
J^\mu = \sqrt{-g}nu^\mu ,  \quad |J|=\sqrt{-g_{\mu\nu}J^\mu J^\nu}, \quad n=\frac{|J|}{\sqrt{-g}}, \quad u^\mu u_\mu=-1\,,
\end{equation}
where $u^{\mu}$ is the 4-velocity of a fluid element. We can vary the Lagrangian in Eq.~(\ref{act}) with respect to the dynamical variables $g_{\m}, J^{\mu},s, \theta,\varphi, \beta_{A}, \alpha^{A}$ to get the dynamical equations for the fluid.
	
The energy-momentum tensor for the matter part can be obtained by varying the action with respect to $g^{\m}$ as:
\begin{eqnarray}
T_{\mu \nu} \equiv -\dfrac{2}{\sqrt{-g}}\dfrac{\delta S}{\delta g^{\mu \nu}}\,.
\label{gentmn}
\end{eqnarray}
Using this definition of the energy-momentum tensor for the case of the perfect fluid action yields 
\begin{eqnarray}
T_{\mu \nu}^{(M)}= \rho u_{\mu } u_{\nu} + \left(n \dfrac{\p \rho}{\p n} -\rho  \right) (u_{\mu } u_{\nu} + g_{\m})\,.
\end{eqnarray}
Comparing with the energy-momentum tensor of standard fluid $T_{\mu \nu}=(\rho + P)u_\mu u_\nu + Pg_{\mu \nu}\,,$ one can identify energy density to be $\rho$ and pressure of the fluid as,
\begin{equation}\label{relativistic fluid pressure}
P= \left(n \dfrac{\p \rho}{\p n} -\rho  \right) \,.
\end{equation}
The pressure and energy density expressions specify a relativistic fluid. Next we will figure out the other energy-momentum tensors and equations of motion coming out from our original Lagrangian.

The Lagrangian for the $k$-essence field is $\mathcal{L}(\phi, X)$. The variation of the field action and  action for the non-minimal coupling  with respect to $\phi$ and $X$ for homogeneous and isotropic FLRW background
\begin{eqnarray}
ds^2 = -dt^2 +a(t)^2 d{\bf x}^2\,,
\end{eqnarray}  
gives the modified field equation as:
\begin{equation}\label{field equation of kessence}
{\mathcal L}_{,\phi} + \nabla_\mu ({\mathcal L}_{,X} \nabla^\mu \phi)
+ f_{,\phi} + \nabla_\mu (f_{,X} \nabla^\mu \phi)=0\,.
\end{equation}
Here $\nabla_\mu A^\nu$ is the covariant derivative of the vector field $A^\nu$.
Commas followed by $\phi$ or $X$ in the subscript implies partial derivatives with respect to $\phi$ or $X$. The last expression gives the field equation of modified  $k$-essence in the presence of mixed non-minimal coupling. To find the energy-momentum tensor of $k$-essence scalar field we vary the field action with $g_{\m}$ and use Eq.~(\ref{gentmn}) to get
\begin{eqnarray}
T_{\mu \nu}^{(\phi)} = - {\mathcal L}_{,X}\,(\partial_\mu \phi)(\partial_\nu \phi)-g_{\mu \nu}\,{\mathcal L}\,.
\label{tphi}  
\end{eqnarray}
Comparing the above energy-momentum tensor with the energy-momentum tensor for a perfect fluid yields the expressions of energy density and pressure of $k$-essence field as:
\begin{equation}\label{}
\rho_{\phi} = \mathcal{L} - 2X\mathcal{L}_{,X} \quad \text{and} \quad P_{\phi} = - \mathcal{L}\,.
\end{equation}
These expressions specify the energy density and pressure of the energy-momentum tensor coming from the $k-$essence sector. 

Varying the part of action responsible for the non-minimal interaction with respect to $g_{\m}$ gives the energy-momentum tensor corresponding to the non-minimal interaction.  Using Eq.~(\ref{gentmn}) we get
\begin{equation}\label{}
T_{\mu \nu}^{({\rm int})} = n\dfrac{\p f}{\p n} u_{\mu} u_\nu +\left(n\dfrac{\p f}{\p n} -f \right)  g_{\m} - f_{,X}(\p_\mu \phi)(\p_\nu \phi)\,.
\end{equation}
Comparing the above expression with the energy-momentum tensor for a perfect fluid we can write the energy density and pressure of this interaction term as: 
\begin{equation}\label{}
\rho_{\rm int} =f-2X f_{,X}\,,  \quad P_{\rm int} = \left(n\dfrac{\p f}{\p n} -f \right)\,. 
\end{equation}
Once we have written the separate energy-momentum tensors we can write the total energy-momentum tensor for the physical system as
\begin{equation}\label{}
T_{\mu \nu}=T_{\mu \nu}^{(\phi)} + T_{\mu \nu}^{(M)} + T_{\mu \nu}^{({\rm int})}\,,
\end{equation}
where $T_{\mu \nu}^{(M)},\,  T_{\mu \nu}^{(\phi)}$ were defined previously.

Variation of the Lagrange multipliers in the Lagrangian in Eq.~(\ref{act}) gives,
\begin{align}
  J^{\mu}:& \qquad	u_{\mu}(\mu_{\rm int} + \mu) +\varphi_{,\mu}+s\theta_{,\mu}+\beta_A\alpha^A_{,\mu} = 0
\label{jeqn}\\ 
s:& \qquad -\left[\frac{\partial \rho}{\partial s}+\frac{\partial f}{\partial s}\right]+nu_{\mu}\theta_{,\mu}=0 \\  
\varphi:& \qquad J^\mu{}_{,\mu}=0 \,,\label{eq:07}\\
\theta:& \qquad (sJ^\mu)_{,\mu}=0 \,,\label{eq:08}\\
\beta_A:& \qquad J^\mu\alpha^A_{,\mu}=0 \,,\label{eq:09}\\
\alpha^A:& \qquad (\beta_AJ^\mu)_{,\mu}=0 \,,\label{eq:10}
\end{align}
where $\mu$ is the chemical potential given by $\mu = {\p \rho}/{\p
  n}$ and $\mu_{\rm int} = {\p f}/{\p n}$. Here $\mu_{\rm int}$ is a new variable defined for our case and it is not the standard chemical potential as
$\mu=(\rho+P)/n$. Eqns.~\eqref{eq:07} and \eqref{eq:08} stand for the
particle number conservation constraint and the entropy exchange
constraint, respectively. Both of these can be written as,
\begin{equation}
\nb_{\mu }(n u^{\mu}) = 0, \quad {\rm and} \quad  \nb_{\mu}(n s u^{\mu}) = 0\,,
\label{eq:constraint relation}
\end{equation}
where $\nb_{\mu }$ is covariant derivative with respect to $g_{\m}$.
\section{Conservation of energy momentum tensor}

Due to the presence of the non-minimal coupling term the individual energy-momentum tensors in general are not conserved separately but the total energy-momentum tensor is conserved. To verify the total energy-momentum conservation we redefine the energy-momentum tensors in a different way: 
$T'_{\m} = T^{(\phi)}_{\mu \nu} - f_{,X}(\p_\mu \phi)(\p_\nu \phi)$ and $\tilde{T}_{\mu \nu} = T^{(M)}_{\m} + T^{({\rm int})}_{\m} + f_{,X}(\p_\mu \phi)(\p_\nu \phi)$. From the redefinition of the energy-momentum tensors we see that $\ti{\rho} = \rho + f$ and $\ti{P} = P + P_{\rm int} $. The Einstein equation in our case will simply be \begin{equation}\label{}
G_{\m} =\kappa^2 (T'_{\mu\nu}+\tilde{T}_{\mu\nu})\,,
\end{equation}
where $\kappa^2 = 8\pi G$. Using the $k-$essence field equation in Eq.~\eqref{field equation of kessence} we have 
\begin{eqnarray}
\nabla^{\mu} T^{\prime}_{\mu\nu} &=& f_{,\phi}\partial_{\nu}\phi - f_{,X}(\partial_{\mu}\phi) \nabla^{\mu} \nabla_{\nu} \phi \equiv  Q_{\nu}\,.
\label{Conserved scalar field}
\end{eqnarray}
Here $Q_\nu$ is just a 4-vector which is defined by the above equation.  
We provide a detail calculation related to energy-momentum conservation in  the appendix appearing in section~\ref{sec:6}. Next we calculate the covariant derivative of $\tilde{T}_{\mu\nu}$. To carry on with this calculation we split the covariant derivative expression into
parallel and perpendicular components to the fluid flow as:
\begin{equation}\label{Fluid emt def}
\nb^{\mu}	\ti{T}_{\m} = h_{\nu}^{\lambda}\nb^{\mu} \ti{T}_{\mu \lambda} - u_\nu u^\lambda \nb^{\mu}T_{\mu \lambda}\,, 
\end{equation}
where $h_{\m} = g_{\m} + u_{\mu}u_{\nu }$. Using Eqs.\eqref{eq:07}-\eqref{eq:10} we can show that 
\begin{equation}\label{perpendicular component}
u_\nu \nb_{\mu}\ti{T}^{\,\m} = -u^{\mu}\dfrac{\p \ti{\rho}}{\p \phi}\nb_{\mu} \phi - u^{\mu}\dfrac{\p \ti{\rho}}{\p X}\nb_{\mu} X\,,
\end{equation}
\begin{equation}\label{parallel component}
h_{\m}\nb_{\lambda}\ti{T}^{\lambda \nu}	 = -h^{\lambda}_{\mu}\left( \dfrac{\p\ti{\rho}}{\p \phi}\nb_{\lambda}\phi + \dfrac{\p\ti{\rho}}{\p X}\nb_{\lambda}X\right)\,. 
\end{equation}
Inserting these results in Eq.~\eqref{Fluid emt def} and we obtain  
\begin{equation}\label{final fluid conservation}
\nb_{\mu}	\ti{T}^{\m} =-\dfrac{\p \ti{\rho}}{\p \phi}\nb^{\nu}\phi  - \dfrac{\p \ti{\rho}}{\p X}\nb^{\nu}X\,.
\end{equation}
In the above equation the partial derivative of $\ti{\rho}$ is with respect to $\phi$ and $X$ while $\ti{\rho} = \rho(n,s) + f(n,s,\phi,X)$. As a result of this fact we see that the last equation becomes 
\begin{equation}\label{}
\nb_{\mu}	\ti{T}^{\m} =-\dfrac{\p f}{\p \phi}\nb^{\nu}\phi  - \dfrac{\p f}{\p X}\nb^{\nu}X = -Q^{\nu}\,,
\end{equation}
showing that the total energy-momentum tensor $T'_{\mu\nu}+\tilde{T}_{\mu\nu}$ is conserved.
\section{Interacting field-fluid theory in the background of FLRW cosmology}
\label{sec:3}	
	
In this paper we have studied a field-fluid non-minimal coupling scenario, where the field part is governed by k-essence scalar field and the fluid part describes a  pressure-less dust type dark matter candidate.  In this section we specify the Friedmann equations and the other necessary details regarding the model we are considering. From Eqs.~(\ref{eq:constraint relation}) we can write down the evolution equations for number density ($n$) and entropy ($s$) of the universe as:
\begin{eqnarray}
\dot{n}+3Hn = 0,\quad  \dot{s}=0\,.
\end{eqnarray}
In the above equation the Hubble parameter is $H=\dot{a}/a$. In the presence of field-fluid interaction Friedmann equations can be written as,
\begin{eqnarray}
  3H^2 &=& \kappa^2 \left(\rho+\rho_{\phi}+\rho_{\rm int}\right)\,,
\label{frd1}\\
2\dot{H}+3H^2 &=& -\kappa^2 \left(P+P_{\phi}+P_{\rm int}\right)\,.
\label{eq:F1}
\end{eqnarray}
Here we have studied the $k$-essence Lagrangian of the form  ${\mathcal L}=-V(\phi)F(X)$ where $F(X)$ is some yet unspecified function of $X$ and $V(\phi)$ is purely a function of $\phi$. The energy density and pressure of the $k$-essence sector can be written as:
\begin{eqnarray}
\rho_{\phi} &=& V(\phi)(2XF_{,X}-F)\,,\quad P_{\phi}=V(\phi)F(X)\,.
\label{eq:F2}
\end{eqnarray}
In purely $k-$essence theories one defines the equation of state (EOS) parameter ($\omega_\phi$) and adiabatic sound speed ($c_{s(\phi)}^2$) as:
\begin{eqnarray}
\omega_\phi = \frac{P_{\phi}}{\rho_{\phi}}\,,\quad c_{s(\phi)}^2 = \frac{dP_\phi/dX}{d\rho_\phi/dX}\,.
\label{eq:F3}
\end{eqnarray}
In defining the sound speed we have followed the convention set in \cite{Garriga:1999vw, Scherrer:2004au}. It was previously noted that this definition of sound speed works with cosmological perturbations. Later on most authors have taken this form of sound speed in $k-$essence theories as a standard definition. Depending on the values of the EOS parameter we can predict whether the universe is accelerating or not. The sound speed parameter plays an important role in studying stability ($c_{s(\phi)}^2>0$) and causality ($c_{s(\phi)}^2<1$) of the theory. Later we will define the sound speed of the field-fluid system in a similar way. We will generalize the definition of the sound speed in the $k-$essence sector in presence of non-minimal coupling.
	
The Scalar field equation appearing in Eq.~\eqref{field equation of kessence} in the context of FLRW universe is given as:
\begin{multline}
\label{field kessence}
\left[ \mathcal{L}_{,\phi} + f_{,\phi}\right] - 3H\dot{\phi}\left[  \mathcal{L}_{,X} + f_{,X} \right] +  \dfrac{\p }{\p X} (P_{\rm int} +f ) (3H \dot{ \phi})\, -
\ddot{ \phi} \left[(\mathcal{L}_{,X} + f_{,X}) + 2X (\mathcal{L}_{,XX} + f_{,XX}) \right]  \\
-\dot{ \phi}^2 (\mathcal{L}_{,\phi X} + f_{,\phi X }) = 0\,.
\end{multline}
Here subscript $XX$ or $\phi X$ implies double partial derivative with respect to $X$ or mixed derivative with respect to $\phi$ and $X$.
After setting up the basic equations governing the field-fluid model we will now study the cosmological system using the dynamical system technique. This technique will help us to understand the dynamical stability of our proposed field-fluid model. 
\section{Dynamical systems approach}
\label{sec:4}	
	
In this section we discuss the dynamical system's techniques in the context of a non-minimally coupled $k$-essence scalar field and a perfect fluid in the cosmological background. We will see that our system can be described by  an autonomous system of ordinary differential equations with three variables. The autonomous equations will be of the form:
\begin{eqnarray}
x^{\prime} = f_1(x,y,z)\,, \quad y^{\prime} = f_2(x,y,z)\,, \quad z^{\prime} = f_3(x,y,z)\,.
\nonumber
\end{eqnarray}
Here $x,y,z$ are three dynamical variables of the system and prime signifies their differentiation with respect to the time parameter. $f_i(x,y,z)$ for $i=1,2,3$ only depend on $x,y,z$ and does not have any explicit dependence on time. We have found the fixed (critical or stationary) points of the system by solving the equations
\begin{eqnarray}
f_i(x_0,y_0,z_0)=0\,,\quad (i=1,2,3)
\nonumber
\end{eqnarray}
where $(x_0,y_0,z_0)$ designates the fixed points of the system. To check the stability of the stationary points we will use linear stability analysis. In this  method we will Taylor expand $f_i(x, y, z)$ around the fixed point $(x_0 , y_0 , z_0)$.  To do a linear stability analysis around the fixed points we require to know the Jacobian matrix of the system. We can define the Jacobian matrix elements, which consists of all the first derivatives of the Taylor series, as:
\begin{eqnarray}
\mathcal{J}_{ij} = \frac{\partial f_i}{\partial x_j}\,.
\end{eqnarray}
Depending on the nature of eigenvalues of the Jacobian matrix evaluated at the critical point we can infer upon the type of fixed points. In the present case as because ${\mathcal J}$ is a $3\times3$ matrix we obtain three eigenvalues from the Jacobian matrix. In a 3-D system if all the real parts of eigenvalues of the Jacobian matrix, at a specific fixed point, have a negative sign we call that fixed point a stable fixed point or stable node. If the sign of the real part of the eigenvalues are positive, at some fixed point, for all eigenvalues then the fixed point becomes an unstable fixed point or unstable node. If at least one of the eigenvalues has an opposite sign the critical point is a saddle point. But when any one of the eigenvalues becomes zero at the fixed point then the linear stability theory fails. 
\subsection{Cosmological dynamics of the non-minimally coupled system}
	
In this subsection we will formulate the nonlinear system of equations, which guide our field-fluid model, in the cosmological background. To write down the equations we first define some dimensionless variables which will be used to set up the dynamical equations of the system. These variables are:
\begin{eqnarray}
x = \dot{\phi}\,, \quad y = \frac{\kappa\sqrt{V(\phi)}}{\sqrt{3}H}\,, \quad \sigma = \frac{\kappa\sqrt{\rho}}{\sqrt{3}H}\,, \quad z = \frac{\kappa^2 f}{3H^2}\,.
\label{eq:D0}
\end{eqnarray}
We have used $\kappa $ and $H$ to convert all the variables into dimensionless quantities. To construct the autonomous system we have also defined some other variables which are dependent on the non-minimal interaction term. These quantities are: 	
\begin{eqnarray}
B =  \dfrac{f_{,\phi}k^2}{H^3},\quad C= \dfrac{\kappa^2 P_{\rm int}}{3H^2},\quad D = \dfrac{\kappa^2 }{3H^2}f_{,X},\quad E=\frac{\kappa^2}{H^3}\dfrac{\p^2 f}{\p \phi \p X  }\,.
\label{eq:D1}
\end{eqnarray}
We can convert Eq.~\eqref{eq:F1} in terms of the chosen variables and express the fourth dynamical variable in terms of these ones. The constraint equation is
\begin{eqnarray}
\sigma^2 =  1- y^2\left(3x^4/4-x^2/2\right) -z +  x^2 D\,.
\label{eq:dyn1}
\end{eqnarray}
The other Friedmann equation, using Eqs.~(\ref{frd1}) and (\ref{eq:F1}), can be written as: 
\begin{eqnarray}
\dfrac{\dot{H}}{H^2} = -\frac32 [\omega \sigma^2 + y^2 F+C+1]\,.	\label{eq:dyn2}		
\end{eqnarray}
Using this information we write down the dynamical equations of the system using the $k-$essence field equation in Eq.~\eqref{field kessence} as:
\begin{eqnarray}
x'=  \dot{x}/H &=& \dfrac{(B/3 + \sqrt{3} \lambda y^3 F) + 3x \left( y^2 F_{,X} + C_{,X}\right)  - x^2(E/3 + \sqrt{3} \lambda y^3 F_{,X})}{[(D - y^2 F_{,X}) + x^2(D_{,X} - y^2 F_{,XX})]} \,,  \label{eq:dynx}\\
y^{\prime}=\dot{y}/H &=& -\dfrac{\sqrt{3}\lambda y^2 x }{2} + \dfrac{3}{2}y  \left[\omega \sigma^2 + y^2 F+C+1 \right]\,,  \label{eq:dyny}\\
z^{\prime}=\dot{z}/H &=&  \left[-3(C+z) + \frac{B}{3}x +D x\, x'  \right]  +      3z\left[\omega \sigma^2 + y^2 F+C+1 \right]\,. \label{eq:dynz}
\end{eqnarray}
Here prime denotes the derivative of the dynamical variables $x,y,z$ with respect to $H dt$. We have used a new variable
$$\lambda = -\dfrac{V_{,\phi}}{\kappa V^{3/2}}\,,$$
to write the above equations. This involves the scalar field potential and its derivative with respect to the field. 
\subsection{Study of the models in the context of dynamical system}

After setting up the autonomous system in  Eqs.~\eqref{eq:dynx}-\eqref{eq:dynz} we will introduce two different interacting models as given in Table~[\ref{table:M0}]. Both of these models require a power law potential of the form
\begin{eqnarray}
V(\phi) = \dfrac{\delta^2}{\kappa^2\phi^2}\,,
\label{plpot}
\end{eqnarray}
where $\delta$ is a dimensionless parameter. This form of the $k-$essence potential has been used previously in various models \cite{ArmendarizPicon:2000ah, Yang:2010vv}. This form of the potential is useful in studying accelerating late phase of the universe using $k-$essence theories. The models of non-minimal field-fluid coupling which we present in this paper are perhaps the simplest ones which give interesting results and moreover the models are chosen in such a way that the dimension of the phase space remains relatively small.  To solve the actual dynamics of the system we have to choose some form of the non-minimal coupling term $f(n,s,\phi, X)$. We assume $f$ depends on $\rho(n,s)$ and some functions of $\phi$ and $X$ separately. The field dependence of $f$ comes from the function $\gamma(\phi)$ and the $X$ dependence of $f$ is obtained from a separate function $g(X)$. To study further we have taken the function $F(X)$ appearing in the $k-$essence Lagrangian as
\begin{eqnarray}
F(X) = X^2 -X\,.
\label{Fform}  
\end{eqnarray}
A similar/same form of $F(X)$ has been used previously in various works on $k-$essence theories \cite{ArmendarizPicon:1999rj,Chiba:1999ka, Scherrer:2004au} in tackling the problems related to the early phase of the universe or late-time accelerated phase of the universe. This form of $F(X)$ is simple and has an extremum for positive values of $X$.
The model parameters $\alpha, \beta, \epsilon $ are all dimensionless quantities. Due to the particular form of the dynamical equations dictating the phase space behavior of the physical system the constant do not appear explicitly in any calculation. The results we present here are practically independent of the non-zero values of $\alpha$. 
\begin{table}[h!]
\centering
\begin{tabular}{|p{1.1cm}| p{2cm} |p{1cm}|p{0.8cm}|p{1.55cm}|p{0.8cm}|p{1.2cm}|p{1cm}|p{1.1cm}|p{1.8cm}|}
\hline 
\multicolumn{10}{|c|}{ $V(\phi) = \dfrac{\delta^2}{\kappa^2\phi^2}$ and $ F(X) = X^2 -X $ } \\
\hline
Models & $ f $ & $\gamma(\phi)$ & $ g(X) $  & B & D & $D_X$ & C & $C_{,X}$ & E\\
\hline
&&&&&&&&&\\
I & $\rho^\epsilon \gamma(\phi) g(X)$ &$ \alpha \frac{\phi}{\kappa}  $ &$\beta X$ & $ \dfrac{3\sqrt{3}\, z\, y}{\delta  }  $ & $ \dfrac{2z}{x^2} $ & 0 & $ z[\epsilon(\omega +1)-1] $ &$D[\epsilon(\omega +1)-1]  $ &$ \dfrac{6\sqrt{3}zy}{x^2 \delta } $\\
\hline
&&&&&&&&&\\
II & $\rho\gamma(\phi) g(X)$ & $\alpha \left( \frac{\phi}{\kappa}\right)^m  $& $ \beta X^n $& $ \dfrac{3\sqrt{3}\, z\, y\, m}{\delta  }  $  &  $ \dfrac{2n\,z}{x^2} $ & $\frac{4n(n-1)z}{x^4} $ & $\omega \, z$ &$\omega D$ &$\dfrac{6\sqrt{3} mnzy}{x^2 \delta } $\\
\hline   
\end{tabular}
\caption{Models chosen for studying the dynamics of non-minimally interacting $k$-essence field with a perfect relativistic fluid.}
\label{table:M0}
\end{table}	
Next we present some details about the separate models shown in the above table. 
\subsection{Model-I}

In this subsection we will analyze the dynamical system described by Eqs.~ \eqref{eq:dynx}-\eqref{eq:dynz} for the choice of functions in Model I. The Friedmann equation in Eq.~\eqref{eq:dyn2}  
\begin{eqnarray}
\dfrac{\dot{H}}{H^2} & = -\dfrac32 \left[ \omega \sigma^2 + y^2 \left(\fx \right)  +z(\epsilon(\omega +1)-1)+1\right]\,, 
\end{eqnarray}
can be solved at the fixed point $(x_{0},y_{0},z_{0})$ for the scale-factor $a$ to give
\begin{equation}\label{scale factor }
a \propto (t-t_{0})^{2/3[\omega \sigma_{0}^2 +y_{0}^2(x_{0}^4/4-x_{0}^2/2) +z_{0}(\epsilon(\omega +1)-1) +1 ]}\,,
\end{equation}
for our specific choice of functions. Here $\sigma_{0}, z_0$ is evaluated at the fixed point and $t_0$ is an integration constant. We can define total energy density and Pressure for this system,
\begin{eqnarray}
\rho_{\rm tot} &=& \rho + \rho_{\phi} +\rho_{\rm int} =\rho + V(\phi) (2XF_{,X}-F) + f-2Xf_{,X}\,,   \\
P_{\rm tot} &=& P + P_{\phi} + P_{\rm int} = P + V(\phi)F + \left(n\dfrac{\p f}{\p n} -f \right)\,. 
\end{eqnarray}
Using the above expressions we can define the total equation of state and adiabatic sound speed as:
\begin{eqnarray}
  \omega_{\rm tot}=\frac{P_{\rm tot}}{\rho_{\rm tot}} &=& {\omega \sigma^2 +y^2 \left(\frac14 x^4-\frac12 x^2\right) +z [\epsilon(\omega +1)-1]}\,,
\label{omega1}\\
c_s^2 &=& \dfrac{dP_{\rm tot}/dX}{d\rho_{\rm tot}/dX} =  \dfrac{\epsilon(\omega+1)D+y^2 F_{,X} }{y^2(F_{,X} + 2XF_{,XX}) - D -2XD_{,X}}\,. 
\end{eqnarray}
Here we have used the definition of the sound speed as given in Eq.~(\ref{eq:F3}). We have generalized the definition of the sound speed in the $k-$essence sector and use it to find the sound speed in presence of  non-minimal coupling. Henceforth we will always use this definition of sound speed. For the fluid part we have assumed the presence of a pressure-less fluid for which $\omega = P/\rho = 0$. In Model I we can find the critical points of the autonomous system described in Eqs.~\eqref{eq:dynx}-\eqref{eq:dynz} by using 
$ x'=0,\, y'=0,\,z'=0$. The critical points  are listed in Table~[\ref{critical point: Model I}]. The scale-factor, as given in Eq.~\eqref{scale factor }, corresponding to this critical point can be written as
\begin{equation}\label{}
a \propto (t-t_0)^{2/3(1+ \omega_{\rm tot})}\,.
\end{equation}
In writing the above expression we have used Eq.~\eqref{omega1}.
In the chosen model there are nine critical points out of which six depends on the value of $\delta$, two depends on the values of $\delta$ and $\epsilon$ and the rest has a constant value. We have calculated $\omega_{\rm tot}$ and $c_s^2$ for these points and inferred upon the accelerated solutions and the causality condition. The points about the accelerated solutions and causality conditions are given in Table~[\ref{Model 01 : omega and sound speed}]. The stability of these points is explicitly shown in Table~[\ref{Model 01 : Stability}]. We have also checked the critical points at infinity and found none. 
\begin{table}[t!]
\centering
\begin{tabular}{|p{1cm}|p{2.7cm}|p{7.5cm}|p{1.7cm}|}
\hline
\multicolumn{4}{|c|}{For $ \omega = 0,  $ The critical points are }\\ 
\hline
Points &	x & y & z \\
\hline
\hline
&	&&\\
$ A $&	$ 0 $ & $ 0 $  & -1 \\
\hline 
&&&\\
$ B_{\mp}$& $ \mp \sqrt{2} $ & $\mp \frac{1}{2} \sqrt{\frac{3}{2}} \delta  $ & 	0 \\
\hline 
&&&\\
$ C_{\mp} $ & $ \mp\frac{\sqrt{\frac{6 \delta ^2-\sqrt{12 \delta ^2+9}+3}{\delta ^2}}}{\sqrt{6}} $  & $\pm  \frac{\sqrt{\frac{12 \delta ^2-2 \sqrt{12 \delta ^2+9}+6}{\delta ^2}} \left(\left(\sqrt{12 \delta ^2+9}+5\right) \delta ^2+\sqrt{12 \delta ^2+9}+3\right)}{6 \delta ^3+4 \delta } $ & 0\\
\hline
&&&\\
$ D_{\mp} $ & $ \mp \frac{\sqrt{\frac{6 \delta ^2+\sqrt{12 \delta ^2+9}+3}{\delta ^2}}}{\sqrt{6}}$ & $ \mp \frac{\sqrt{2} \sqrt{\frac{6 \delta ^2+\sqrt{12 \delta ^2+9}+3}{\delta ^2}} \left(\left(\sqrt{12 \delta ^2+9}-5\right) \delta ^2+\sqrt{12 \delta ^2+9}-3\right)}{6 \delta ^3+4 \delta } $ & 0\\
\hline
&&&\\	
$ E_{\mp} $ & $ \mp \frac{\sqrt{2 \delta ^2 \epsilon ^3+20 \epsilon -24}}{\sqrt{\delta ^2 \epsilon ^2 (3 \epsilon -2)}}$ & $ \mp \dfrac{\delta  \epsilon  \sqrt{\delta ^2 \epsilon ^2 (3 \epsilon -2)}}{\sqrt{6 \delta ^2 \epsilon ^3+60 \epsilon -72}} $ & $ \frac{\epsilon  \left(\delta ^2 \epsilon +6\right)-12}{9 \epsilon -6} $ \\
\hline
\end{tabular}
\caption{Critical Points for Model I. }
\label{critical point: Model I}
\end{table}
\begin{table}[h!]
\centering
\begin{tabular}{|p{1cm}|p{8cm}|p{4.3cm}|}
\hline
Points & \quad $\omega_{\rm tot}$ & $ c_{s}^2  $ \\
\hline 
$ A $ & $ 1-\epsilon $ & $ -\epsilon $ \\
\hline 
$ B_{\mp} $ & $ 0 $ & $ 1/5 $ \\
\hline 
&&\\
$ C_{\mp} $ & $ \dfrac{1}{24 \delta ^4 \left(6 \delta ^3+4 \delta \right)^2} \left(12 \delta ^2-2 \sqrt{12 \delta ^2+9}+6\right) \newline \left(6 \delta ^2-\sqrt{12 \delta ^2+9}+3\right) \left(\frac{6 \delta ^2-\sqrt{12 \delta ^2+9}+3}{6 \delta ^2}-2\right) \newline   \left(\left(\sqrt{12 \delta ^2+9}+5\right) \delta ^2+\sqrt{12 \delta ^2+9}+3\right)^2 $   & $ \dfrac{\dfrac{6 \delta ^2-\sqrt{12 \delta ^2+9}+3}{6 \delta ^2}-1}{\dfrac{6 \delta ^2-\sqrt{12 \delta ^2+9}+3}{2 \delta ^2}-1} $ \\
\cline{2-3}
&&\\
& $\omega_{tot} <0$, $\delta \ne 0 $ & Not satisfied \\
\hline 
&&\\
$ D_{\mp} $ & $ \dfrac{1}{12 \delta ^4 \left(6 \delta ^3+4 \delta \right)^2} \left(6 \delta ^2+\sqrt{12 \delta ^2+9}+3\right)^2  \newline \left(\left(\sqrt{12 \delta ^2+9}-5\right) \delta ^2+\sqrt{12 \delta ^2+9}-3\right)^2 \newline \left(\dfrac{6 \delta ^2+\sqrt{12 \delta ^2+9}+3}{6 \delta ^2}-2\right) $ & $ \dfrac{\dfrac{6 \delta ^2+\sqrt{12 \delta ^2+9}+3}{6 \delta ^2}-1}{\dfrac{6 \delta ^2+\sqrt{12 \delta ^2+9}+3}{2 \delta ^2}-1} $  \\
\cline{2-3}
& $\omega_{tot} <0$, $ \delta \lessgtr \mp \dfrac{2}{\sqrt{3}} $ for any $ \epsilon $  &  Satisfied $\forall$ $\delta \in \mathbb{R} $ for any $ \epsilon $\\
\hline 
$ E_{\mp} $ &$ \dfrac{\delta ^2 \epsilon ^2 \left(2 \delta ^2 \epsilon ^3+20 \epsilon -24\right) \left(\frac{2 \delta ^2 \epsilon ^3+20 \epsilon -24}{\delta ^2 \epsilon ^2 (3 \epsilon -2)}-2\right)}{4 \left(6 \delta ^2 \epsilon ^3+60 \epsilon -72\right)} \newline +\dfrac{(\epsilon -1) \left(\epsilon  \left(\delta ^2 \epsilon +6\right)-12\right)}{9 \epsilon -6} $  & Both $\epsilon$ and $\delta$ dependent  \\
\cline{2-3}
&$\omega_{tot}  < 0$, For $\epsilon < 3/2$ $\forall$ $ \delta \in \mathbb{R} $ & For $\epsilon = 1$, $\delta \gtrless \pm 2.31 $ \newline $\epsilon =0,2,3$, $\delta \in \mathbb{R}$. \\
\cline{2-3}
& $\omega_{tot} >0$ ,  For $\epsilon > 3/2$ $\forall$ $ \delta \in \mathbb{R} $ & For $\epsilon = -1$, $-2.83 <\delta < 2.83$, \newline
$ \epsilon = -2 $, Not Satisfied \newline 
$\epsilon = -3$, $\delta \gtrless 3.27$ \\
\hline
\end{tabular}
\caption{Study of Effective $ \omega_{\rm tot} $ and sound speed condition $ 0< c_s^2 <1 $ for Model I.}
\label{Model 01 : omega and sound speed} 
\end{table}

Next we discuss the nature of the critical points from the inputs in the various tables.
\begin{enumerate}
\item Point $A$: This fixed point is independent of any model parameter.  The eigenvalues for the Jacobian matrix is undefined at this point and we cannot infer properly on the stability of this critical point by linearizing around this point. As a consequence of this we have not included this point in our phase space analysis.
		
\item Points $ B_{\pm} $: These critical points depend on only $\delta$
and is well behaved for any $\delta \in \mathbb{R}$. The cosmological system is matter dominated near these points as $\omega_{\rm tot} =0$ at these points. These points are stable for $\epsilon>3/2$ and in some specific range in $\delta$ given in Table~[\ref{Model 01 : Stability}]. In all of the cases we discuss here it will turn out that the EOS and the effective sound speed remains the same for the pair of points $B_\pm$ or $C_\pm$ or other conjugate pairs of critical points.
		
\item Points $ C_{\pm} $: These critical points exists for $\delta \ne 0, \delta \in \mathbb{R}$. These points give rise to accelerating solutions for $\delta \ne 0$, but the adiabatic sound speed limit is not satisfied near these points.  These points can be stable fixed points or saddle points depending on the values of $\epsilon$ and $\delta$. To have more knowledge about  stability of this point one can see the contents in Table~[\ref{Model 01 : Stability}]. 
		
\item Points $ D_{\pm} $: These critical points depend on $\delta$ and have real values for $ \delta \ne 0, \delta \in \mathbb{R} $. Near these points we have accelerating solutions for any value of $\epsilon$ and $\delta \gtrless \pm \dfrac{2}{\sqrt{3}}$. Here $\delta \gtrless \pm \dfrac{2}{\sqrt{3}}$ stands for
$\delta > \dfrac{2}{\sqrt{3}}$ or $\delta < -\dfrac{2}{\sqrt{3}}$.
The sound speed condition is also satisfied for $ \delta \in \mathbb{R}$ for any $\epsilon$. These are stable points for some specific ranges in $\epsilon$ and $\delta$ given in Table~[\ref{Model 01 : Stability}]. 
		
\item Points $ E_{\pm} $: These critical points depend on $\epsilon$ and $ \delta $. Near these points we can have both accelerating and non-accelerating solutions depending on the values of $\epsilon$ as shown in Table~[\ref{Model 01 : omega and sound speed}]. These critical points are real for $ \epsilon > 6/5 $, $ \delta  \ne 0$, and for $ \frac{2}{3}<\epsilon \leq \frac{6}{5} $, $ \delta \gtrless \pm \sqrt{2} \sqrt{-\frac{5 \epsilon -6}{\epsilon ^3}} $. The stability condition given in Table~[\ref{Model 01 : Stability}] shows that these points can be saddle points or stable points depending on the values of the parameters $\epsilon$ and $\delta$.
\end{enumerate}  
\begin{table}[h!]
\begin{tabular} {|p{1cm} |p{3cm}| p{6cm}| p{4cm}| }
\hline
Points & $\epsilon$ & $\delta$ & Stability \\
\hline 
$ A  $ & any & any & Undefined \\
\hline 
$ B_{\mp} $ & $\epsilon > 3/2$ & $ -1.15<\delta \leq -0.71 $,\newline  $ 0.71 \le \delta \le 1.15 $ & Stable \\
\cline{2-4} 
& any & $ \delta  \ne 0 $, and 	not in above range & Saddle \\
\hline 
$ C_{\mp} $ & $ \epsilon >-\dfrac{3}{\delta ^2}-1.73 \sqrt{\frac{4 \delta ^2+3}{\delta ^4}} $ & $\delta \ne 0$, $\delta \in \mathbb{R}$ & Stable \\
\cline{2-4}
& $\epsilon > -7.6$ & $\delta = 1$ & Stable \\
& $\epsilon > -2.64$  & $ \delta = 2  $ & Stable
\\
& 	$\epsilon > -1.54$ & $ \delta = 3  $ & Stable
\\
& &  & Otherwise Saddle \\
\hline 
$ D_{\mp} $ & $ 0<\epsilon \leq \frac{3}{2} $ &$  \delta <-\sqrt{\frac{12-6 \epsilon }{\epsilon ^2}} $ , $ \delta >\sqrt{\frac{12-6 \epsilon }{\epsilon ^2}} $ & Stable \\
\cline{2-4}
& $ \epsilon >\frac{3}{2} $ & $ \delta \lessgtr\mp1.155 $ & Stable \\
\cline{2-4}
& 1& $ \gtrless \pm2.45 $ & Stable \\
&  &  & Otherwise  Saddle \\
\hline 
$ E_{\mp} $ & 1 & 1&  Saddle  \\ 
\cline{2-4}
& 1 & 2 & Stable\\ 
\cline{2-4}
& 1 & 3 & Saddle \\
\cline{2-4}
& 2 & 1 & Saddle \\
\cline{2-4}
& 2 & 2 & Saddle \\
\cline{2-4}
& 2 & 3 & Saddle \\
\cline{2-4}
& -1 & 1& Saddle \\
\cline{2-4}
& -1 & 2& Saddle \\
\cline{2-4}		
& -2 & 1 & Saddle\\
\cline{2-4}	
& -2 & 3& Stable \\
\hline 
\end{tabular}
\caption{Stability Condition of Model I for $\omega = 0$.}
\label{Model 01 : Stability}
\end{table}

\begin{figure}[t!]
\begin{minipage}[b]{0.5\linewidth}
\centering
\includegraphics[scale=.6]{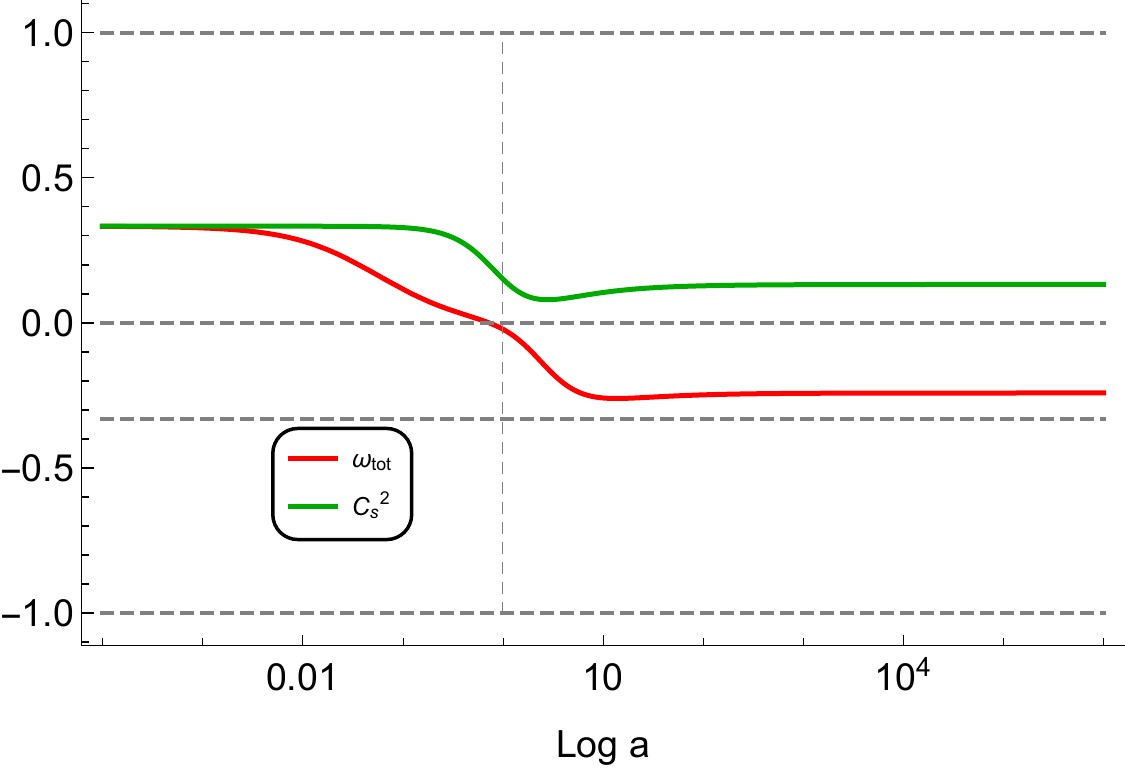}
\caption{Evolution of $\omega_{\rm tot}$ and $c_s^2$ for the case $\epsilon = 1,\, \delta =2,\, \omega = 0$} 
\label{fig:evo_01}
\end{minipage}
\hspace{0.2cm}
\begin{minipage}[b]{0.5\linewidth}
\centering
\includegraphics[scale=.6]{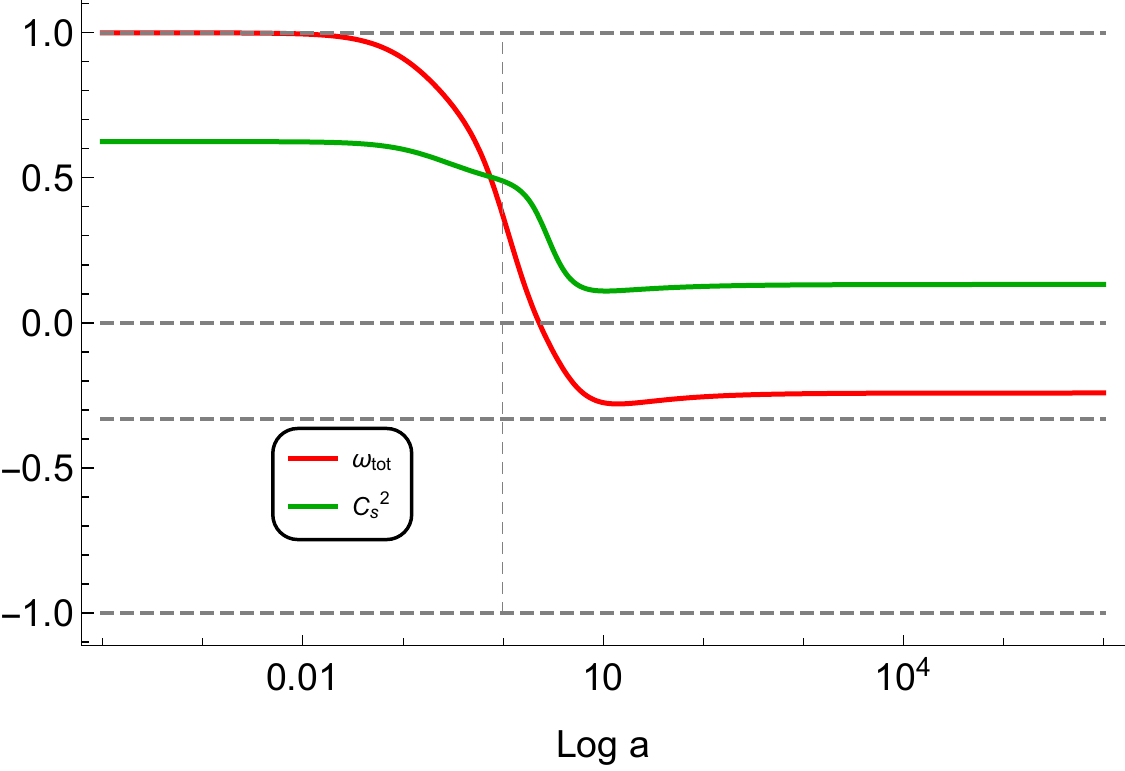}
\caption{Evolution of $\omega_{\rm tot}$ and $c_s^2$ for the case $\epsilon = 2,\, \delta =2,\, \omega = 0$}
\label{fig:evo_02}
\end{minipage}
\end{figure}

From our model study it is seen that non-minimal coupling of $k-$essence field and dust solution can give rise to late time cosmic acceleration. Next we will like to analyze the trajectories in the phase space of Model I. The difficulty for such an analysis arises from the fact that the dynamical variables $(x,y,z)$ are defined in the intervals $ -\infty<x<\infty $, $ -\infty<y<\infty $, $-\infty<z<\infty$ and hence we require to scan an infinite volume. In short, phase space is non-compact. To overcome this difficulty we first try to produce a finite phase space volume by using the transformations: 
\begin{equation}\label{}
X = \arctan x, \quad  Y =\arctan y, \quad \; Z= \arctan z\,.
\end{equation}
With the choice of these variables our phase space is now finite. Henceforth we will work with the variable $X,Y,Z$.
\begin{itemize}	
\item {\textbf{Case I: ($\epsilon =1$)}} \; We have plotted few trajectories for this case as shown in Fig.~[\ref{fig:phase_space_01}]. In Fig.~[\ref{fig:phase_space_01}], where we have used $\epsilon=1$, one can  see a similar behavior of trajectories along the $X-$axis except for the position of points, which get flipped off. The trajectories starts at some point, which is difficult to locate in this redefined phase space, and evolve towards the points $B_{\pm}$, $C_{\pm}$, $D_{\pm}$ and $E_{\pm}$. The trajectories that evolve toward $B_{\pm}$ (red curves) gets repelled as these points are saddle points and ultimately reach $E_\pm$, which are attractor points. The effective EOS near $B_\pm$ is 0 showing that near these points the system is in a matter dominated phase. The universe near $E_\pm$ shows accelerated expansion but these points may not be physically viable as the sound speed limits are violated near them. If we use $\delta=3$ the points $E_\pm$ becomes acceptable points as the sound speed limits remain within the bounds in that case. Our analysis shows that non-minimal coupling can produce various phases of the universe. The fate of the phase trajectories moving towards $D_\pm$ (black curves) is similar to the previous case as $D_\pm$ is also a saddle point. We get accelerated expansion near $D_\pm$. The figure also shows that some trajectories move towards $C_\pm$, which are attractors. Near $C_\pm$ we get accelerated expansion but the EOS near these points give -2.8. Non-minimal interaction produces phantom like behavior near these points.  
\begin{figure}[t!]
\begin{minipage}[b]{0.5\linewidth}
\centering
\includegraphics[scale=.6]{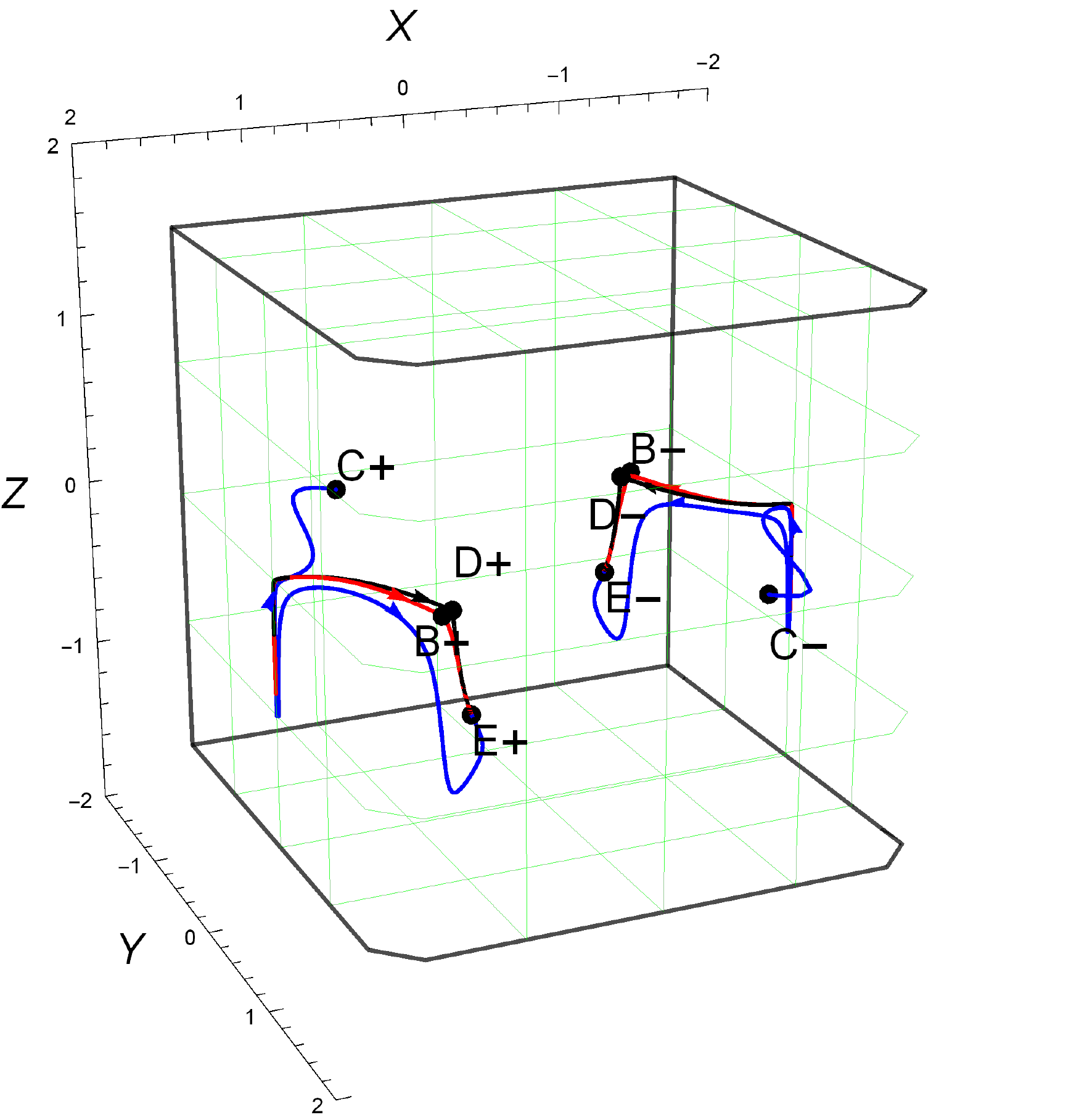}
\caption{System trajectories in redefined phase space for $\epsilon = 1,\, \delta =2,\, \omega = 0$.} 
\label{fig:phase_space_01}
\end{minipage}
\hspace{0.2cm}
\begin{minipage}[b]{0.5\linewidth}
\centering
\includegraphics[scale=.6]{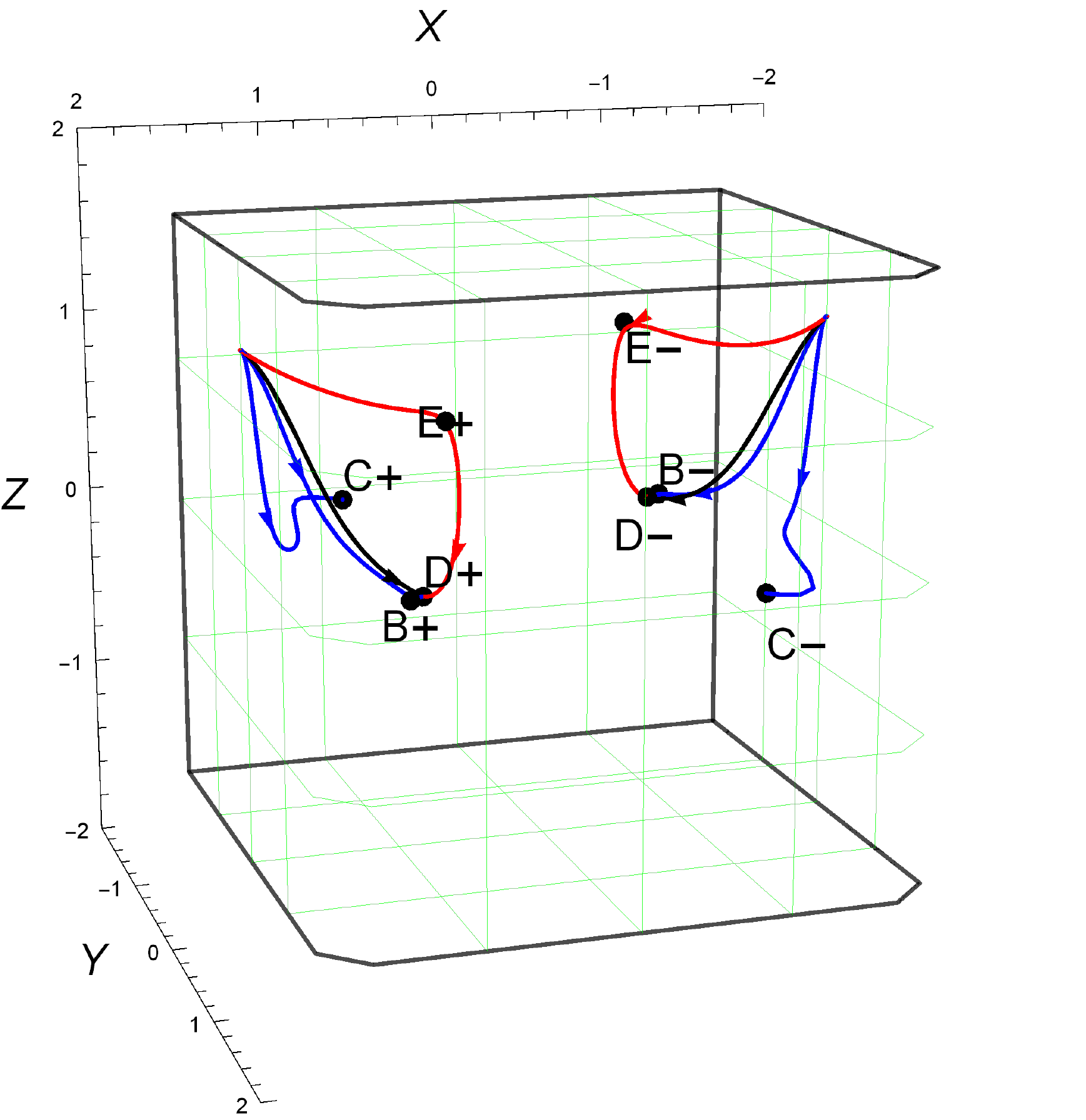}
\caption{System trajectories in redefined phase space for $\epsilon = 2,\, \delta =2,\, \omega = 0$.}
\label{fig:phase_space_02}
\end{minipage}
\end{figure}

\item {\textbf{Case II: ($\epsilon = 2 $) }}\;  In this case we study the dynamics of the system when $\epsilon = 2$ and $\delta =2$ when the fluid satisfies $\omega =0$. In Fig.~[\ref{fig:phase_space_02}] the redefined phase space is drawn for such a system where we plot few trajectories.  These trajectories evolves towards the points $B_{\pm}$, $C_{\pm}$, $D_{\pm}$ and $E_\pm$.
The trajectories which move towards $D_{\pm}$ are first attracted towards the points $E_{\pm}$ (red curves) or $B_\pm$ (blue curves). The cosmological system is in a radiation dominated phase near $E_\pm$ where the effective EOS is 1/3.
Near $B_\pm$ the system shows properties of matter domination. The set of points $E_\pm$ and $B_\pm$ are saddle points. Some trajectories (black curves) directly move towards the points $D_\pm$ showing that these points are attractors. The EOS near $D_\pm$ gives -0.24 showing that near these points we get accelerated expansion. Some trajectories are attracted towards the points $C_\pm$, which are attractors. The cosmological system near these points is dominated by phantom like matter and sound speed limits are violated near these points.
\end{itemize}  
The discussion on the dynamical evolution of the system following Model I parameters show that we can have cosmological evolution from radiation domination to late time dark energy dominated phase in presence of non-minimal interaction of $k-$essence field with a pressure-less fluid. The features become apparent from  
the case where $\epsilon =2$. To visualize the evolution of the total EOS and sound speed variation through the various phases we have also plotted the evolution of $\omega_{\rm tot}$ and $c_{s}^2$ against the logarithm of the
scale-factor in Fig.~[\ref{fig:evo_01}]  and Fig.~[\ref{fig:evo_02}] for one particular trajectory of the phase space. In Fig.~[\ref{fig:evo_01}] we explicitly see how a radiation dominated era evolves to the phase of late time acceleration. Throughout the evolution the sound speed limit remains within permissible range. In Fig.~[\ref{fig:evo_02}] the possible early time EOS corresponds to stiff matter. Then the dynamical system reaches the radiation dominated phase for some time and ultimately reaches the accelerated expansion phase  near the points $D_{\pm}$.
\subsection{Model-II}
	
Next we investigate the dynamical stability of Model II where the interaction term
is of the form
$$f = \rho \alpha \left(\dfrac{\phi}{\kappa}\right)^m \beta X^n\,. $$
Using Eq.~\eqref{eq:dyn2} we can express the scale-factor $a$ in terms of cosmic time $t$ at the fixed point $(x_{0},y_{0},z_{0})$ as:
\begin{eqnarray}
a \propto (t-t_{0})^{2/3[\omega \sigma_{0}^2 +y_{0}^2(x_{0}^4/4-x_{0}^2/2) +z_{0}\omega]}\,.
\label{eq:M2-a}
\end{eqnarray}  
For this model we have calculated total energy density, total pressure, EOS and adiabatic sound speed. Total energy density and pressure for this model are given as:
\begin{eqnarray}
\rho_{\rm tot} &=& \rho + V(\phi) (2XF_{,X}-F) + f-2Xf_{,X}\,,   \nonumber\\
P_{\rm tot} &=& P + V(\phi)F + \left(n\dfrac{\p f}{\p n} -f \right)\,. 
\label{eq:M2-b}
\end{eqnarray}
From the above expressions we can write down the total EOS and adiabatic sound speed as:
\begin{eqnarray}
\omega_{\rm tot}	&=& \omega z + y^2\left(\frac14 x^4-\frac12 x^2\right)\,, \nonumber\\
c_s^2 &=& \dfrac{(\omega+1)D+y^2 F_{,X} }{y^2(F_{,X} + 2XF_{,XX}) - D -2XD_{,X}}\,. 
\label{eq:M2-c}
\end{eqnarray}
If the hydrodynamic fluid is pressure-less then the grand EOS is independent of the interaction term and sound speed is dependent on the model parameters $\delta,m,n$. In the present case the dynamical system has $8$ critical points which are listed in Table~[\ref{Tab:M2a}]. The critical points, $A_{\pm}$, $C_\pm$ and $D_\pm$, depend only on  $\delta$. The critical points $B_{\pm}$ depends on all  parameters $\delta, m, n$. In the present case the EOS parameter and sound speed limits put some restrictions on the value of model parameters $\delta,m,n$. In Table~[\ref{Tab:M2b}] and [\ref{Tab:M2c}], we have shown these constraints. Depending on the values of model parameters $\delta,m,n$ we have listed below the nature of the critical points in a tabular manner in Table~[\ref{Tab:M2d}].	  
\begin{table}[h!]
\centering
\begin{tabular}{|p{1cm}|p{4cm}|p{7.5cm}|p{2.8cm}|}
\hline
\multicolumn{4}{|c|}{For $ \omega = 0,  $ The critical points are }\\ 
\hline
Points &	x & y & z \\
\hline
\hline 
&&&\\
$ A_{\pm }$& $ \pm \sqrt{2} $ & $ \pm \frac{1}{2} \sqrt{\frac{3}{2}} \delta  $ & 	0 \\
\hline 
&	&&\\
$ B_{\pm }$& $\pm \frac{\sqrt{2 \delta^2-\frac{4}{3} m (m+2)}}{\delta }	$   & $\pm \frac{3 \delta ^2}{(m+2) \sqrt{6 \delta ^2-4 m (m+2)}}$  & $\frac{3 \delta ^2-2 m (2 m+5)-4}{(m+2)^2 (2 n-1)}$ \\
\hline 
&&&\\
$ C_{\pm} $ & $ \pm \frac{\sqrt{\frac{6 \delta ^2-\sqrt{12 \delta ^2+9}+3}{\delta ^2}}}{\sqrt{6}} $  & $ \mp \frac{\sqrt{\frac{12 \delta ^2-2 \sqrt{12 \delta ^2+9}+6}{\delta ^2}} \left(\left(\sqrt{12 \delta ^2+9}+5\right) \delta ^2+\sqrt{12 \delta ^2+9}+3\right)}{6 \delta ^3+4 \delta } $ & 0\\
\hline
&&&\\
$ D_{\pm} $ & $ \pm\frac{\sqrt{\frac{6 \delta ^2+\sqrt{12 \delta ^2+9}+3}{\delta ^2}}}{\sqrt{6}} $ & $ \pm\frac{\sqrt{2} \sqrt{\frac{6 \delta ^2+\sqrt{12 \delta ^2+9}+3}{\delta ^2}} \left(\left(\sqrt{12 \delta ^2+9}-5\right) \delta ^2+\sqrt{12 \delta ^2+9}-3\right)}{6 \delta ^3+4 \delta } $ & 0\\
\hline
\end{tabular}
\caption{Sets of critical Points for Model II}
\label{Tab:M2a}
\end{table}
\begin{table}[h!]
\centering
\begin{tabular}{|p{1.5cm}|p{1cm}|p{3cm}|p{2cm}|p{2.5cm}|}
\hline
Critical Points & $n$ & $m$ & $\delta$ & $\omega_{\rm tot.}$  \\
\hline
\hline
$ A_{\pm} $ & Any & Any & Any & $\omega_{\rm tot.} = 0 $ \\
\hline
$ B_{\pm} $ & Any & $ m< -2, m>0 \newline m =0 \newline -2<m\leq 0 $ & $\delta \in \mathbb{R} \newline \delta \in \mathbb{R}\newline \delta \in \mathbb{R}$ &  $ \omega_{\rm tot.}<0 $  \newline $\omega_{\rm tot.}=0$ \newline $\omega_{\rm tot.} > 0 $   \\
\hline 
\hline  
$ C_{\pm} $ & Any & Any & $ \delta \ne 0 $ & $\omega_{\rm tot.} < 0 $ \\
\hline 
$ D_{\pm} $ & Any & Any & $ \delta \gtrless \pm \frac{2}{\sqrt{3}} $ & $\omega_{\rm tot.} < 0 $ \\
\hline 
\end{tabular}
\caption{Restrictions on model parameters $\delta,m,n$ from EOS}  
\label{Tab:M2b}
\end{table}

We will now discuss the nature of these critical points in a detailed way.
\begin{enumerate}
\item Points $A_{\pm}$:  These points exist when  $\delta \neq 0$. This condition is obtained from the existence criterion of the critical points. We always get matter dominated era ($\omega_{\rm tot}=0$) at this point.  The EOS parameter does not show any kind of dependency on model parameter $\delta,m,n$.  Adiabatic sound speed squared always lies between 0 to 1 at these points. For some choice of $m$ and $\delta$ we see that the critical points may behave as saddle points whereas for other values of the parameters these points remain stable points. 
		
\item Points $B_{\pm}$: These points exist when  $\delta^2 > \frac23 m(m+2)$ and $\delta \neq 0$. This condition is obtained from the existence criterion of the critical points. We get an accelerating expansion solution for $m<-2 ~{\rm and}~ m>0$ when  $\delta$ is real.  Other than these we have found $\omega_{\rm tot}\geq0$. The EOS parameter does not show any kind of dependency on model parameter $n$. In Table~[\ref{Tab:M2c}] we write down the values of model parameters $\delta,m,n$ for which adiabatic sound speed squared lies between 0 to 1. For some choice of $m$ and $n$ we see that the critical points are stable and for other choices the points behave as saddle points. 		
		
\item Points $C_{\pm}$: These fixed points exist for $\delta \neq 0$.  We get accelerating cosmological expansion for all the real values of $\delta$, except $\delta \neq 0$ . Adiabatic sound speed always lies outside of 0 to 1 for $\delta \in \mathbb{R}$. As these points are independent of $n$ so the stability of these points only depends on the model parameters $\delta,m$.
		
\item Points $D_{\pm}$: These points exist when $\delta \neq 0$.  In this case also we get an accelerating universe solution for $ \delta \gtrless \pm \frac{2}{\sqrt{3}} $. Adiabatic sound speed also lies between 0 to 1 when $\delta \neq 0$. Like $C_{\pm}$ these points are also independent of $n$. Stability of these points depend on the values of model parameters $\delta,m$. We have shown the stability conditions of this model in Table~\ref{Tab:M2d}.
\end{enumerate}

\begin{table}[h!]
\centering
\begin{tabular}{|p{1.5cm}|p{1cm}|p{1cm}|p{7cm}|p{2cm}|}
\hline
Critical Points & $n$ & $m$ & $\delta$ &  $0<c_s^2<1$ \\
\hline
\hline
$ A_{\pm} $ & Any & Any & Any & Satisfied\\
\hline
& 0 & $ 0 $ & $\delta \in \mathbb{R}\;  \& \;  \delta \ne 0 $ & Satisfied      \\
			
& 0 & $ 1 $ & $\delta \in \mathbb{R}\;  \& \;  \delta \gtrless \pm 2,\, -\sqrt{2}<\delta <\sqrt{2} $ & Satisfied      \\
			
& 0 & $ 2 $ & $\delta \in \mathbb{R}\;   \& \; \delta \gtrless \pm 4 \sqrt{\frac{2}{3}},\,-\frac{4}{\sqrt{3}}<\delta <\frac{4}{\sqrt{3}} $ & Satisfied \\
			
$ B_{\pm} $	 & 1 & $ 1 $	& $ \delta \in \mathbb{R} \; \&  \; \delta \gtrless \pm \frac{4}{ \sqrt{3}} $ & Satisfied \\
			
& 1 & 2 & $ \delta \in \mathbb{R} \; \&  \; \delta \gtrless \pm \frac{4 \sqrt{7}}{3} $ & Satisfied \\
			
& 2 & 1 & $ \delta \in \mathbb{R} \; \&  \;   \mp 3\sqrt{2}  \lessgtr \delta \lessgtr  \mp \frac{6}{\sqrt{7}}$  & Satisfied  \\
			
& 2 & 2 &  $ \delta \in \mathbb{R} \; \&  \; \mp 4\sqrt{\frac{7}{3}} \lessgtr \delta \lessgtr \mp \frac{16}{\sqrt{21} } $ &  Satisfied \\
\hline 
$ C_{\pm} $  & Any& Any & $ \delta \in \mathbb{R} $ & Dissatisfied \\
\hline
$ D_{\pm} $  & Any& Any & $ \delta \ne 0 $ & Satisfied \\
\hline  
\end{tabular}
\caption{Constraints on sound speed from model parameters $\delta,m,n$}
\label{Tab:M2c}
\end{table}
\begin{table}[h!]
\centering 
\begin{tabular}{|p{1.5cm}|p{1cm}|p{3cm}|p{6cm}|p{2cm}|}
\hline
Critical Points & $n$ & $m$ & $\delta$ & Stability\\
\hline
\hline
$ A_{\pm} $ & any & $ m<0, m \in \mathbb{R} $ &  $ -\frac{2}{\sqrt{3}}<\delta \leq -\frac{1}{\sqrt{2}}, \frac{1}{\sqrt{2}}\leq \delta <\frac{2}{\sqrt{3}}  $ & Stable.   \\
\cline{2- 5}
&&&& Otherwise saddle \\
\hline
$ B_{\pm} $	 & 0 & 1 & $ \pm 3 $ & Stable \\
& 1 & 0 & 2 & Saddle. \\
& 2 & 1& 2 & Stable. \\
\hline
$ C_{\pm} $ & any & $m>-0.43 \sqrt{4 \delta ^2+3}-1.25$ & $ \delta \ne 0 $ &  Stable. \\
\cline{2- 5}
&&& & Otherwise Saddle \\
\hline
$ D_{\pm} $ & any & $ m \le 0  $ & $ \delta \gtrless \pm 1.15$ & Stable \\
\cline{2- 5}
& any & $ m > 0 $ & $ \delta <-0.82 \sqrt{2 m^2+5 m+2} \, , $ \newline $ \delta >0.82 \sqrt{2 m^2+5 m+2} $ & Stable \\
\cline{2- 5}
&&&& Otherwise Saddle. \\
\hline
\end{tabular}
\caption{Nature of stability at different values of the model parameters $\delta,m,n$}
\label{Tab:M2d}
\end{table}
Now we will discuss the nature of the critical points and also plot phase space trajectories, as it has done for Model I. Here we will use different combinations of model parameters $m$, $n$ and $\delta$  and see the nature of the redefined phase space trajectories.

\begin{figure}[t!]
\begin{minipage}[b]{0.5\linewidth}
\centering
\includegraphics[scale=.6]{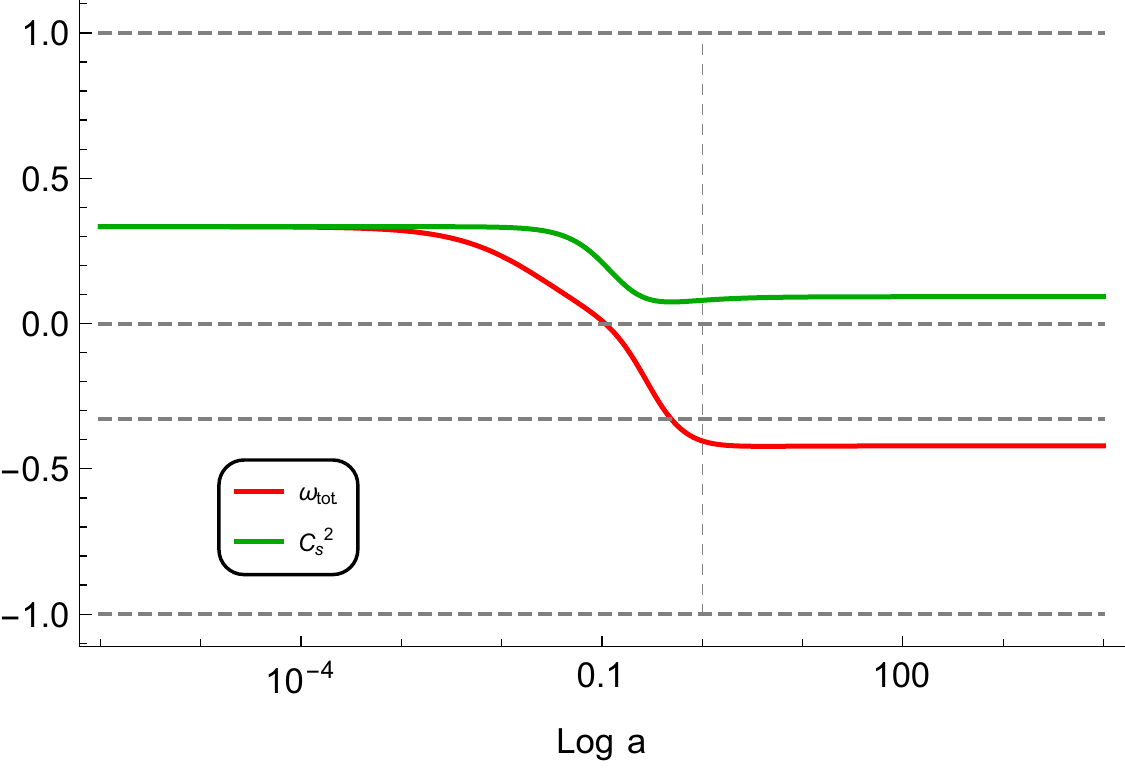}
\caption{Evolution of $\omega_{\rm tot} $ and $ c_{s}^2 $ for $ n=0,  m=1, \delta = 3$.}
\label{fig:M2a2}
\end{minipage}
\hspace{0.2cm}
\begin{minipage}[b]{0.5\linewidth}
\centering
\includegraphics[scale=.6]{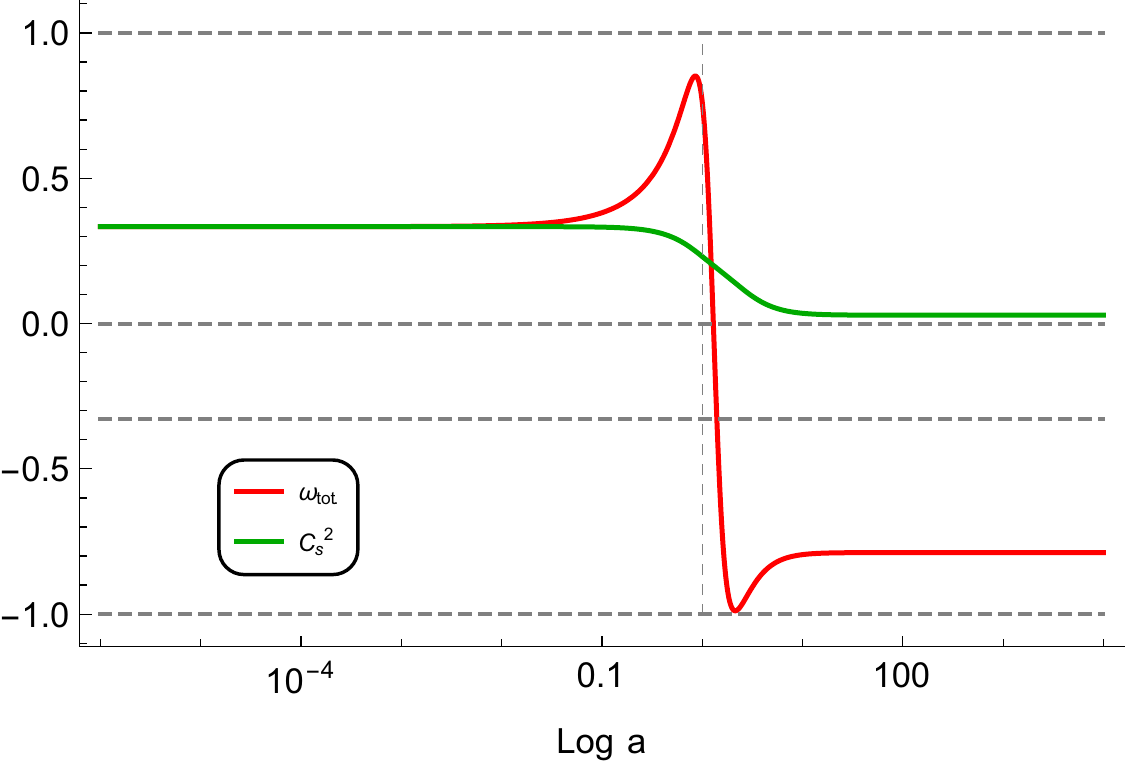}
\caption{Evolution of $\omega_{\rm tot} $ and $ c_{s}^2 $ for  $n=1, m=3,  \delta = 10$.}
\label{fig:M2b2}
\end{minipage}
\end{figure}

\begin{figure}[t!]
	\centering
	\includegraphics[scale=.6]{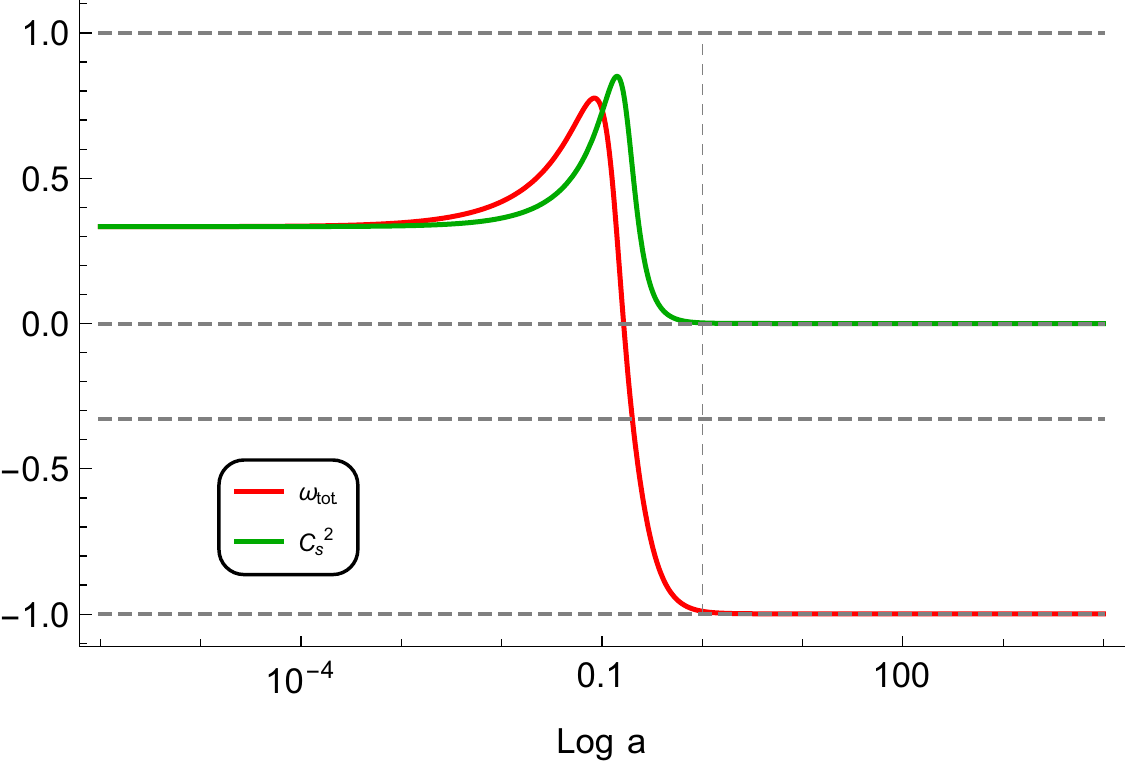}
	\caption{Evolution of $\omega_{\rm tot} $ and $ c_{s}^2 $ corresponds to $n=1,  m=1, \delta = 2000$. The  early phase resembling radiation dominated universe transforms to late time accelerated phase in the dark energy dominated universe.}
\label{fig:M2c2}
\end{figure}

\begin{itemize}
\item {\textbf{{Case I: ($n=0, m=1,\delta = 3$)}}}\; In Fig.~[\ref{fig:M2a1}] we have depicted the phase plot for a universe where $\omega=0$. The $k-$essence scalar field $\phi$ mainly controls the dynamics in this case as the kinetic term, in the non-minimal coupling, is absent since $n=0$. In this plot the trajectories originate from some point and evolve towards $A_{\pm}, B_{\pm},\  C_{\pm} {\rm and} \, D_{\pm}$ points. Few trajectories evolve towards the points $A_{\pm}$ producing matter dominated cosmological scenario with EOS $0$. Near these points the sound speed limits are obeyed.  As because $A_{\pm}$  are saddle points the trajectories are repelled from these points and goes towards stable points $D_{\pm}$. The critical points $B_{\pm}$, near which we get accelerating cosmological expansion with EOS $-0.33$, are also saddle points and consequently the trajectories are repelled from these points and move towards stable points $D_{\pm}$ which are global attractors.  Near $C_{\pm}$ we obtain a phantom type of accelerating universe solutions with EOS $-2.02$ but the difficulty with the fixed points $C_{\pm}$ is that near these points the sound speed limits are violated. These points are stable fixed points. Near the critical points $D_{\pm}$ we have accelerating universe solutions with EOS $-0.42$ and moreover near these points the adiabatic sound speed limits are respected.

\begin{figure}[t!]
\begin{minipage}[b]{0.5\linewidth}
\centering
\includegraphics[scale=.6]{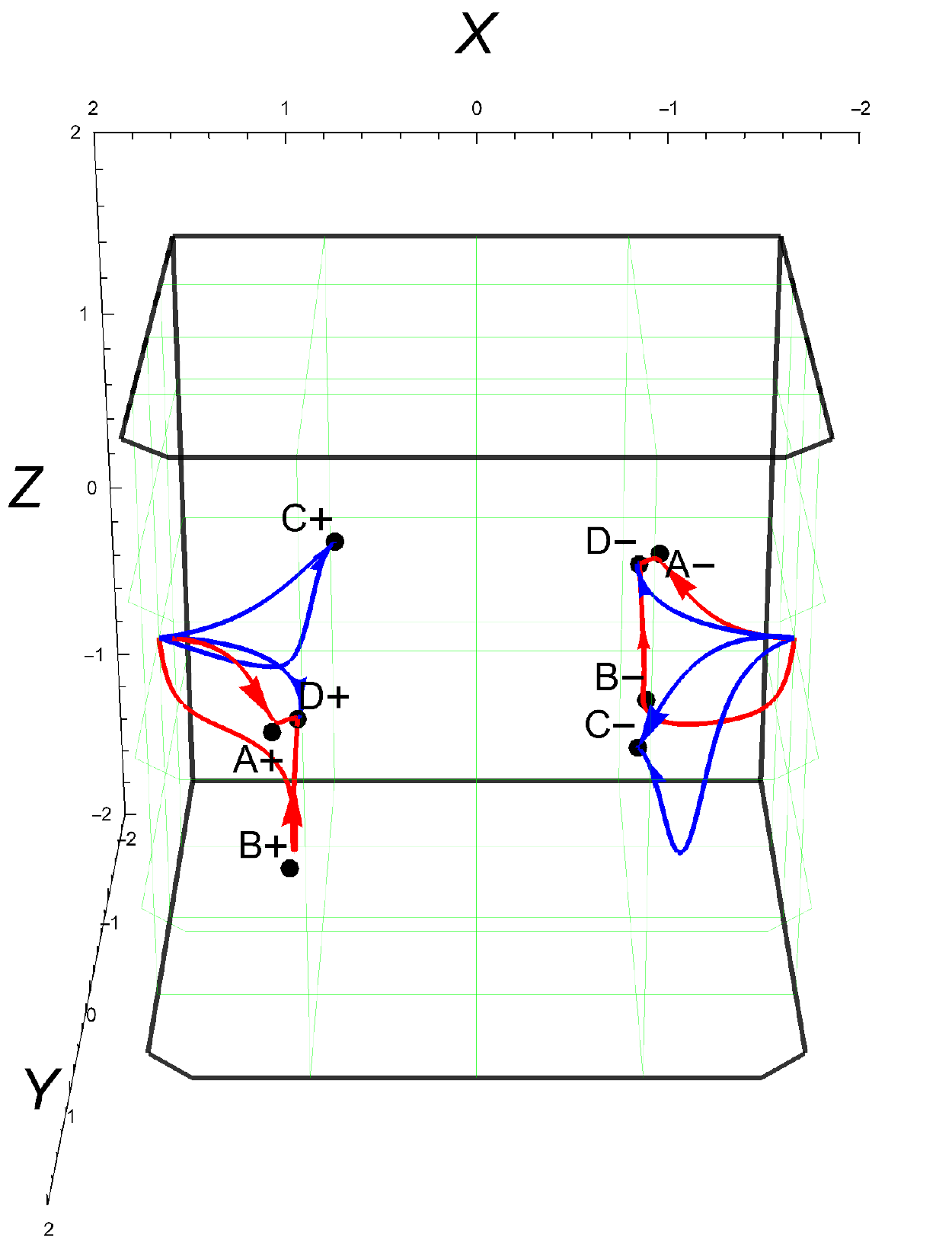} 
\caption{Phase space plot of Model II for $n=0, m=1,  \delta = 3$ } 
\label{fig:M2a1}
\end{minipage}
\hspace{0.2cm}
\begin{minipage}[b]{0.5\linewidth}
\centering
\includegraphics[scale=.7]{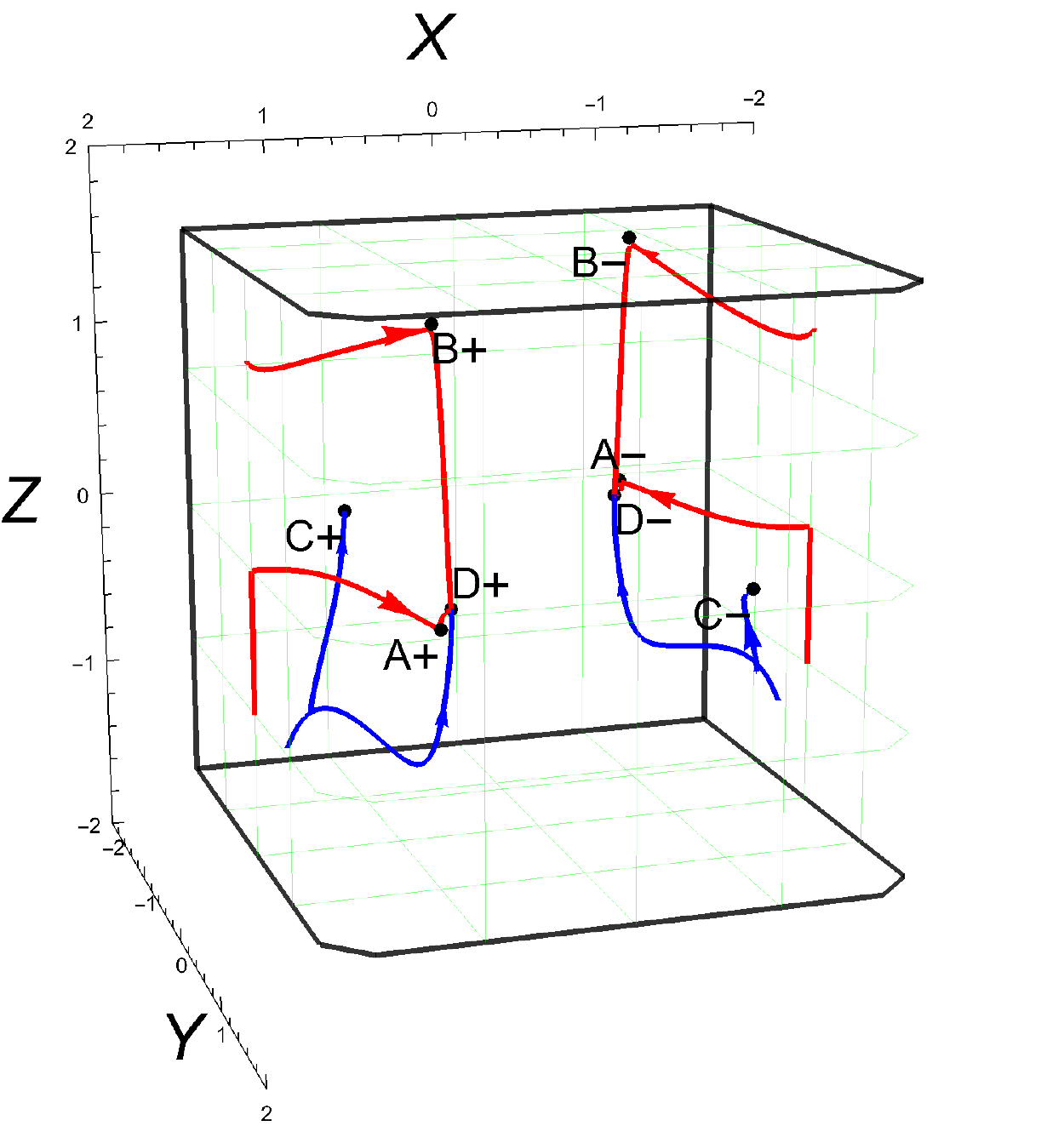}
\caption{System trajectories in redefined phase space for $n=1,  m=3, \delta = 10$}
\label{fig:M2b1}
\end{minipage}
\end{figure}
\item {\textbf{{Case II : ($n=1,  m=3, \delta = 10$)}}}\; In Fig.~[\ref{fig:M2b1}] we have plotted the phase space behavior for the dynamical system for some other choice of parameters. This case corresponds to a universe where the non-minimal coupling term is slightly dominated by the scalar field of the $k$-essence sector due to the unequal values of $n$ and $m$. In this case, points $A_{\pm}$ are saddle points and give a matter-dominated solution with EOS $0$. Similarly, points $B_{\pm}$ also turn out to be saddle points but they give an accelerating solution with EOS $-0.6$.  Trajectories moving towards them ultimately gets repelled and reach the stable points $D_\pm$. Critical points $C_{\pm}$ are stable and EOS around these points is $-1.25$.  Near $C_{\pm}$ points the system shows phantom  behavior and violate adiabatic sound speed condition and as a result  these points are physically not acceptable. In this scenario, $D_{\pm}$ are stable fixed points and act as global attractors. The EOS around these points is $-0.8$ and sound speed is $0.02$. If we choose a high value of model parameter $\delta$, then sound speed comes close to $10^{-3}$ or becomes less than this limit, which matches closely the sound speed limit given in Ref.\cite{Kumar:2019gfl}. From this we can infer that the dynamics of the system near $D_{\pm}$ points can describe the late time phenomenology of our universe. In the next case we present such a case where $\delta$ has a relatively high value.

\begin{figure}[t!]
	\centering
	\includegraphics[scale=.6]{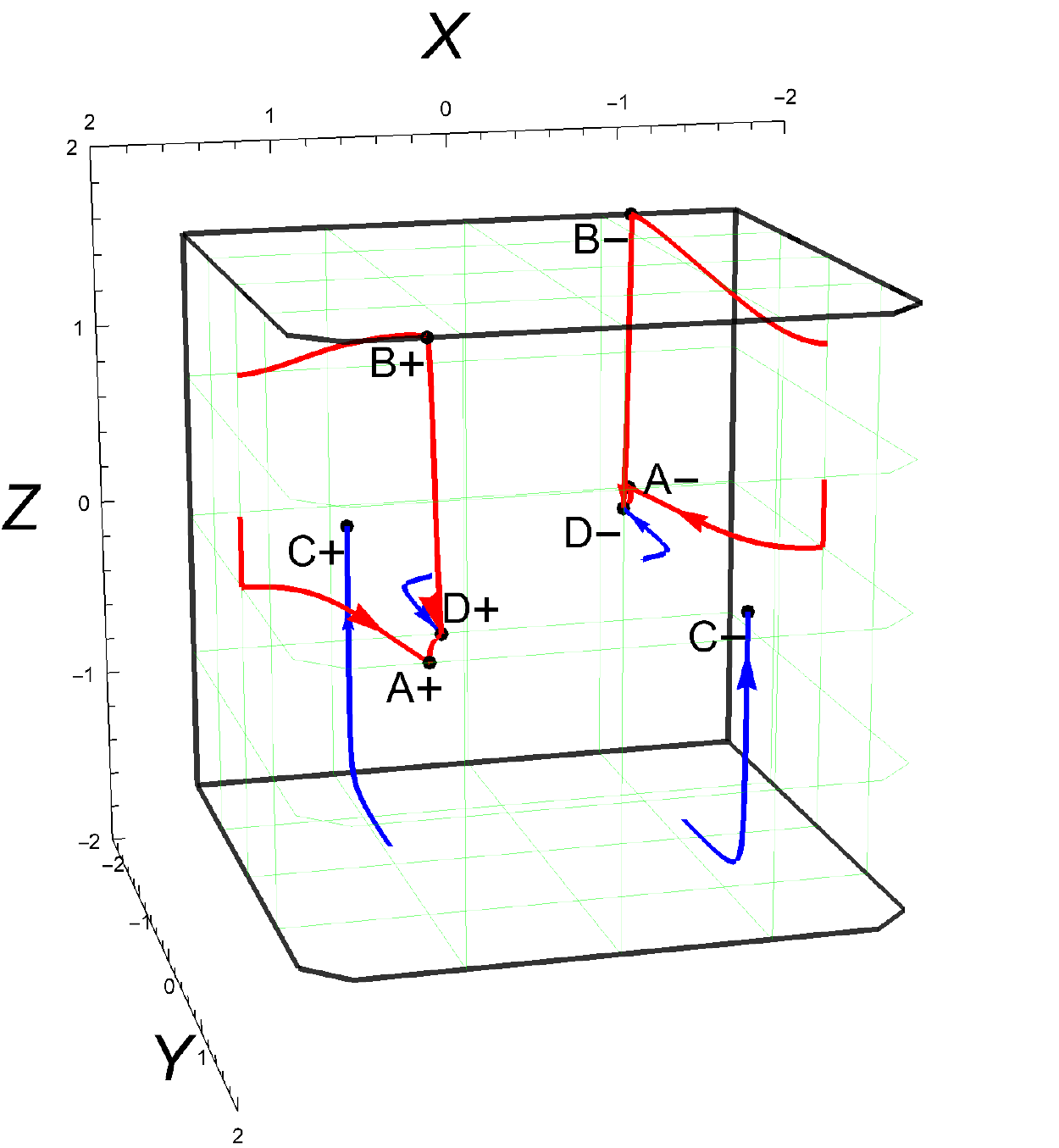}
	\caption{System trajectories in redefined phase space for $n=1,  m=1, \delta = 2000$}
	\label{fig:M2c1}
\end{figure}

\item {\textbf{{Case III : ($n=1,  m=1, \delta = 2000$)}}}\; In Fig.~[\ref{fig:M2c1}] we have shown the phase space plots for case-III. In this case, both the scalar field and kinetic term in the non-minimal field-fluid coupling equally play an important role.  In the present case $A_{\pm}$ and $ B_{\pm} $ are the saddle points, while $C_{\pm},D_{\pm}$ are stable fixed points. Some of the trajectories are attracted towards $C_{\pm}$, with EOS $ -1.001 $. The solutions corresponding to $C_{\pm}$ are attractor points characterizing the accelerating solution, but these solutions are physically not viable because sound speed is negative $ \sim -10^{-4} $ near them. EOS of the field-fluid system near $A_{\pm}$ is $0$ and does not violate the sound speed limits. The points $B_{\pm}$ give an accelerating solution with EOS $-0.333$, and physically acceptable, but it does not produce enough acceleration. The  points $D_{\pm}$ give accelerating solution with EOS $-0.998$, are globally attractor points with sound speed $ \sim 10^{-4} $ \cite{Kumar:2019gfl}. These points produce enough acceleration that is very close to $ -1 $. We may infer that the system's dynamics near these points may describe the late time phenomenology of our universe.
\end{itemize}
The above analysis shows that Model-II can also produce important results that describes the late time behavior of the universe in presence of non-minimal field-fluid coupling. In  Figs.~[\ref{fig:M2a2},\ref{fig:M2b2},\ref{fig:M2c2}] we have shown the evolution of the total EOS parameter ($\omega_{\rm tot}$) and adiabatic sound speed ($c_s^2$) for a particular trajectory of phase space in three different cases mentioned above. Both of these quantities have been plotted against $\log a$ where $a$ is the scale-factor of the FLRW universe.  The plots show the transition of the EOS parameter from positive to negative values as a function of the scale-factor. This suggests that the present cosmological model, which can produce accelerated expansion in the late universe,  can be smoothly matched to a previous radiation dominated phase of the universe.  

\section{Conclusion}
\label{sec:5}

In this article we have studied the cosmological effects of non-minimal interaction between $k$-essence scalar field and a pressure-less fluid using the variational method. In the absence of any non-minimal interaction between these two sectors the pressure-less perfect fluid represents the dark matter sector and the $k-$essence scalar field sector acts primarily as the source of dark energy. In our work both these sectors interact with each other and produce an effective dark sector which can give rise to accelerated expansion of the universe. We have introduced the non-minimal interaction term $f(n,s,\phi, X)$ which depends on the field strength as well as the kinetic term of the $k$-essence sector. Except the $k-$essence part this interaction also depends upon the fluid energy density.

Our study is carried out in the background of an isotropic and homogeneous universe (FLRW universe). We have presented the Friedmann and scalar field equations in our work. Both of these equations have been modified in the presence of non-minimal interaction term. We have systematically derived all the field equations from the first principles and proved the energy-momentum conservation equation of field-fluid interacting theory in section~\ref{sec:2}. In our work the non-minimal interaction term is not given by any fundamental equations but has to be chosen using phenomenological understanding.

We have specified two models and studied the stability of both models in the context of inverse power law $k-$essence potential. The phase spaces for these models are three dimensional and unbounded. We have introduced some new variables to make the phase space compact. For Models I and II the critical points depend on the parameter appearing in the chosen potential, which suggests that the potential plays an essential role in the nature of the outcome. For these models we have found  accelerated expansion solutions in the late time phase of the universe.

In our work the $k-$essence sector is not the only agent which produces the accelerating universe scenario. Traditionally the $k-$essence fields not interacting with any fluid can produce dark energy. In our model due to the non-minimal coupling of the scalar field and the fluid there arises an effective dark sector which causes the accelerating universe scenario. To understand the effect of the dark sector in our case we have defined a grand EOS ($\omega_{\rm tot}$) and an effective sound speed $c_s$. These quantities specify the dark energy sector in presence of field-fluid coupling. The dynamical systems approach used in Model I and Model II shows the existence of various critical points in phase space. Some of these critical points are stable while others turn out to be unstable. It is shown that there exist trajectories in the phase space which designates evolving cosmological scenarios with accelerated expansion. Moreover some of these developments are physically interesting as in some of these cases the sound speed do satisfy the standard limits expected from them. Our work conclusively shows that if the $k-$essence sector indeed has a non-minimal coupling with a pressure-less fluid then also we can have reasonably well behaved accelerating expansion of the universe. Our work relies on the form of the $k-$essence potential and the form of the function $F(X)$ appearing in the $k-$essence Lagrangian. The forms we have chosen are standard ones but in future one can work with other choices.

An interesting feature of our work shows up in the plots of $\omega_{\rm tot}$ and $c_s^2$ where these objects are plotted with respect to logarithm of the scale-factor of the universe. When we plot the above quantities for some trajectory in the phase space we see that some of these trajectories do start from a region where the grand EOS is that of a radiation dominated universe. In our work we have not used radiation fluid but it is interesting to see that non-minimal coupling of $k-$essence with pressure-less fluid can produce a radiation fluid like phase. This is one of the strong features of our work, the non-minimal coupling can effectively produce radiation like EOS as well as stiff EOS. The cosmological consequence of this fact may be far reaching. In this paper we do not present more material related to these topics but in future publications these resemblances will be studied in detail. For the present work we can see that for some cases our model produces a universe which can safely be joined with a proper radiation phase existing in the early universe.   

In the light of non-minimally coupled field-fluid interaction it is seen that we can have interesting cosmological consequences.  As because the traditional dark matter sector is coupled to the $k-$essence field it will be interesting to study the problem of structure formation using our models. The present models do give some important results regarding accelerated expansion of the universe and produces many interesting questions which require further research in the near future.

\paragraph{Acknowledgement}
Authors would like to thank the referee for his/her valuable suggestions. A.C would like to thank Indian Institute of Technology, Kanpur for supporting this work by means of Institute Post-Doctoral Fellowship \textbf{(Ref.No.DF/PDF197/2020-IITK/970)}.

\section{Appendix}
\label{sec:6}
In this section, we will provide an explicit calculation of Eqns.\eqref{Conserved scalar field} and \eqref{final fluid conservation}.
\subsection{Conservation equation of $k$-essence scalar field sector}

We have
\begin{eqnarray}
T^{\prime}_{\mu\nu} &=& -\mathcal{L}_{,X} (\partial_{\mu}\phi)(\partial_{\nu}\phi) - g_{\mu\nu}\mathcal{L} - f_{,X}(\partial_{\mu}\phi)(\partial_{\nu}\phi)\,,\nonumber\\
\nabla^{\mu} T^{\prime}_{\mu\nu} &=& -[\nabla^{\mu} \left(\mathcal{L}_{,X}(\partial_{\mu}\phi)\right)(\partial_{\nu}\phi)+\mathcal{L}_{,X} (\partial_{\mu}\phi) \nabla^{\mu} \nabla_{\nu} \phi + g_{\mu\nu} (\mathcal{L}_{,X}\partial^{\mu}X+\mathcal{L}_{,\phi}\partial^{\mu}\phi) \nonumber\\
&+& \nabla^{\mu} \left(f_{,X}(\partial_{\mu}\phi)\right)(\partial_{\nu}\phi)+f_{,X} (\partial_{\mu}\phi) \nabla^{\mu} \nabla_{\nu} \phi]\,.
\label{eq:Tmunuprime}
\end{eqnarray}
As $X = -\frac12 {(\partial_{\mu}\phi)}{(\partial^{\mu}\phi)}$
\begin{eqnarray}
\partial_{\nu} X &=& -(\partial_{\mu}\phi) (\nabla^{\mu}\nabla_{\nu} \phi)\,.\nonumber
\end{eqnarray}
We can use the above relation in Eq.~\eqref{eq:Tmunuprime} and get:
\begin{eqnarray}
\nabla^{\mu} T^{\prime}_{\mu\nu} &=& -[\nabla^{\mu}\left(\mathcal{L}_{,X} (\partial_{\mu}\phi) +f_{,X}(\partial_{\mu}\phi)\right)(\partial_{\nu}\phi) +\mathcal{L}_{,X} (\partial_{\mu}\phi) \nabla^{\mu} \nabla_{\nu} \phi -\mathcal{L}_{,X} (\partial_{\mu}\phi) \nabla^{\mu} \nabla_{\nu} \phi \nonumber\\
&+& \mathcal{L}_{,\phi}\partial_{\nu}\phi +f_{,X} (\partial_{\mu}\phi) \nabla^{\mu} \nabla_{\nu} \phi]\,. 
\label{eq:Tmunuphi1}
\end{eqnarray}
From the $k-$essence equation of motion in presence of non-minimal coupling
\begin{eqnarray}
\nabla_{\mu}\left(\mathcal{L}_{,X} (\partial^{\mu}\phi) +f_{,X}(\partial^{\mu}\phi)\right) = -(\mathcal{L}_{,\phi}+f_{,\phi})\,,\nonumber
\end{eqnarray}
and the relation in Eq.~\eqref{eq:Tmunuphi1} we get
\begin{eqnarray}
\nabla^{\mu} T^{\prime}_{\mu\nu} &=&  f_{,\phi}\partial_{\nu}\phi - f_{,X}(\partial_{\mu}\phi) \nabla^{\mu} \nabla_{\nu} \phi\,.\nonumber
\end{eqnarray}
We will write the above equation as:
\begin{equation}
\nabla^{\mu} T^{\prime}_{\mu\nu} = f_{,\phi}\partial_{\nu}\phi - f_{,X}(\partial_{\mu}\phi) \nabla^{\mu} \nabla_{\nu} \phi \equiv Q_{\nu}\,,
\label{eq:coupling}
\end{equation}
where $Q_\nu$ is a 4-vector.
\subsection{Conservation equation of fluid sector}

Using the definition
\begin{equation}\label{}
h_{\m} = g_{\m} + u_{\mu}u_{\nu }\,,
\end{equation}
we can write
\begin{equation}\label{Energy Momentum Tensor1}
\nb^{\mu}	\ti{T}_{\m} = h_{\nu}^{\lambda}\nb^{\mu} \ti{T}_{\mu \lambda} - u_\nu u^\lambda \nb^{\mu}\ti{T}_{\mu \lambda}\,. 
\end{equation}
To proceed we calculate the following term as:
\begin{equation}
\begin{split}
u_\nu \nb_{\mu}\ti{T}^{\,\m}  &= u_\nu \nb_{\mu} \left[ (\tilde{\rho} + \tilde{P})u^\mu u^\nu + \tilde{P}g^{\m}\right]\\
& = u_\nu \nb_{\mu}(\ti{\rho}+ \ti{P})u^{\mu}u^{\nu} + u_\nu (\ti{\rho}+ \ti{P}) \left[ (\nb_{\mu}u^{\mu}) u^{\nu}  +u^{\mu} \nb_{\mu}U^{\nu} \right]  + u_{\nu}g^{\m} \nb_{\mu}\ti{P} \\ 
& =- \nb_{\mu}(\ti{\rho}+ \ti{P})u^{\mu} - (\ti{\rho}+ \ti{P})(\nb_{\mu}u^{\mu}) + u_{\nu}(\ti{\rho}+ \ti{P})u^{\mu}\nb_{\mu}u^{\nu} + u^{\mu}\nb_{\mu}\ti{P} \\ 
& =- u^{\mu}\nb_{\mu}\ti{\rho} - (\ti{\rho}+ \ti{P})(\nb_{\mu}u^{\mu}) + (\ti{\rho}+ \ti{P})u_{\nu}u^{\mu}\nb_{\mu}u^{\nu}\\
& = -u^{\mu}\nb_{\mu}\ti{\rho} - \left( \ti{\rho} + \dfrac{\p \ti{\rho}}{\p n} n -\ti{\rho}\right)\nb_{\mu}u^{\mu}  + 0 \\
& = -u^{\mu}\left(\dfrac{\p \ti{\rho}}{\p n}\nb_{\mu}n + \dfrac{\p \ti{\rho}}{\p s}\nb_{\mu}s + \dfrac{\p \ti{\rho}}{\p \phi}\nb_{\mu}\phi + \dfrac{\p \ti{\rho}}{\p X}\nb_{\mu}X \right) - \dfrac{\p \ti{\rho}}{\p n}\left( \nb_{\mu}(nu^{\mu}) - u^{\mu}\nb_{\mu}n \right) \\
& = -u^{\mu}\dfrac{\p \ti{\rho}}{\p \phi}\nb_{\mu} \phi - u^{\mu}\dfrac{\p \ti{\rho}}{\p X}\nb_{\mu} X\,.
\nonumber
\end{split}
\end{equation}
In a compact form we can write the result as:
\begin{equation}\label{perpendicular component1}
u_\nu \nb_{\mu}\ti{T}^{\,\m} = -u^{\mu}\dfrac{\p \ti{\rho}}{\p \phi}\nb_{\mu} \phi - u^{\mu}\dfrac{\p \ti{\rho}}{\p X}\nb_{\mu} X\,.
\end{equation}
In deriving the last equation we have used $\nb_{\mu}(nu^{\mu}) = 0\,,\,\,\nb_{\mu}s = 0$ , $u_{\nu}\nb_{\mu}u^{\nu} = 0  = u^{\nu}\nb_{\mu}u_{\nu}$ and also $u^{\nu}u_{\nu} =-1$. 

Next we calculate $h_{\m}\nb_{\lambda}\ti{T}^{\lambda \nu}$. After some steps we can write
\begin{equation}\label{parallel component1}
h_{\m}\nb_{\lambda}\ti{T}^{\lambda \nu}	 = 2n\,u^{\lambda}\nb_{\big[\lambda}(\ti{\mu}u_{\mu\big]})-h^{\lambda}_{\mu}\left( \dfrac{\p\ti{\rho}}{\p \phi}\nb_{\lambda}\phi + \dfrac{\p\ti{\rho}}{\p X}\nb_{\lambda}X\right)\,, 
\end{equation}
where
$$\tilde{\mu}\equiv\frac{\tilde{\rho}+\tilde{P}}{n}\,,\quad {\rm and} \quad
2\nb_{\big[\lambda}(\ti{\mu}u_{\mu\big]})\equiv \nabla_\lambda(\tilde{\mu}u_\mu)-
\nabla_\mu(\tilde{\mu}u_\lambda)\,.$$
One must note that $\tilde{\mu}=\mu_{\rm int}={\p f}/{\p n}$ introduced in section \ref{sec:2}. To evaluate the first term in right side of the last equation  we are going to use 
Eq.~(\ref{jeqn}) which gives
\begin{equation}\label{}
\ti{\mu}u_{\mu} = -\left( \varphi_{,\mu} +s \theta_{,\mu} +\beta_{A} \alpha_{,\mu}^{A}\right)\,. 
\end{equation}
Next we evaluate the antisymmetric term as:
\begin{equation}\label{}
\begin{split}
u^{\lambda}\nb_{\big[\lambda}(\ti{\mu}u_{\mu\big]})& = u^{\lambda}\nb_{\big[\lambda}\left[ \nb_{\mu\big]}\varphi + s\,\nb_{\mu\big]} \theta + 	\beta_{A}\nb_{\mu\big]} \alpha^{A}			\right]  \\
& = u^{\lambda}\left[\nb_{\big[\lambda} \nb_{\mu\big]}\varphi + s\, \nb_{\big[\lambda}\nb_{\mu\big]}\theta + \nb_{\big[\lambda} \beta_{A} \nb_{\mu\big]} \alpha^{A}\right] \\
& = u^{\lambda}\left[\nb_{\lambda}\beta_{A}\nb_{\mu}\alpha^{A} - \nb_{\mu}\beta_{A}\nb_{\lambda}\alpha^{A} \right] \\
& =0
\end{split}
\end{equation}
since $\nb_{\big[\lambda}\nb_{\mu\big]}g = 2(\nb_{\lambda} \nb_{\mu} - \nb_{\mu} \nb_{\lambda})g = 0$ where $g$ is a scalar function. The covariant derivatives commute for torsion-less space time when applied on scalar functions. Here the fluid variables $\beta_{A} \, , \, \alpha^{A}$ are scalars.

We can use Eq.~\eqref{perpendicular component1} and Eq.~\eqref{parallel component1} in Eq.~\eqref{Energy Momentum Tensor1} as
\begin{eqnarray}\label{}
\nb_{\mu}\ti{T}^{\m} & = & h^{\nu}_{\lambda}\nb_{\mu}\ti{T}^{\mu \lambda} - U^{\nu}U_{\lambda}\nb_{\mu}\ti{T}^{\mu \lambda} \\
& = & -\dfrac{\p \ti{\rho}}{\p \phi}\nb^{\nu}\phi  - \dfrac{\p \ti{\rho}}{\p X}\nb^{\nu}X\,
\end{eqnarray}
Remembering that $\tilde{\rho}=\rho +f$ where $\rho$ does not depend upon $\phi$ or
$X$ we can finally write
\begin{equation}\label{final fluid conservation1}
\nb_{\mu}\ti{T}^{\m} =-\dfrac{\p \ti{f}}{\p \phi}\nb^{\nu}\phi + f_{,X}(\partial_{\mu}\phi) \nabla^{\mu} \nabla_{\nu} \phi = -Q^{\nu}\,.
\end{equation}
This completes the proof of the energy-momentum conservation of our non-minimally interacting field-fluid model.


\end{document}